\documentclass[twocolumn,tighten,twocolappendix]{aastex63}

\usepackage{amsmath}

\newcommand{\change}[1]{\textcolor{blue}{#1}}
\newcommand{\fix}[1]{\textcolor{black}{#1}}

\hypersetup{linkcolor=blue,citecolor=blue,filecolor=blue,urlcolor=blue}
\begin{document}

\title{Limits on Millimeter Continuum Emission from Circumplanetary Material in the DSHARP Disks}

\author[0000-0003-2253-2270]{Sean~M.~Andrews}
\affiliation{Center for Astrophysics \textbar\ Harvard \& Smithsonian, 60 Garden Street, Cambridge, MA 02138, USA}

\author{William Elder}
\affiliation{Center for Astrophysics \textbar\ Harvard \& Smithsonian, 60 Garden Street, Cambridge, MA 02138, USA}

\author[0000-0002-8537-9114]{Shangjia Zhang}
\affiliation{Department of Physics and Astronomy, University of Nevada, Las Vegas, 4505 S. Maryland Parkway, Las Vegas, NV 89154, USA}

\author[0000-0001-6947-6072]{Jane Huang}
\altaffiliation{NASA Hubble Fellowship Program Sagan Fellow}
\affiliation{Department of Astronomy, University of Michigan, 323 West Hall, 1085 South University Avenue, Ann Arbor, MI 48109, USA}

\author[0000-0002-7695-7605]{Myriam Benisty}
\affiliation{Unidad Mixta Internacional Franco-Chilena de Astronom{\'{\i}}a, CNRS/INSU UMI 3386, Departamento de Astronom{\'{\i}}a, Universidad de Chile, Camino El Observatorio 1515, Las Condes, Santiago, Chile}
\affiliation{Univ. Grenoble Alpes, CNRS, IPAG, F-38000 Grenoble, France}

\author[0000-0002-2358-4796]{Nicol{\'a}s T. Kurtovic}
\affiliation{Max-Planck-Institut f{\"u}r Astronomie, K{\"o}nigstuhl 17, 69117 Heidelberg, Germany}

\author[0000-0003-1526-7587]{David J. Wilner}
\affiliation{Center for Astrophysics \textbar\ Harvard \& Smithsonian, 60 Garden Street, Cambridge, MA 02138, USA}

\author[0000-0003-3616-6822]{Zhaohuan Zhu}
\affiliation{Department of Physics and Astronomy, University of Nevada, Las Vegas, 4505 S. Maryland Parkway, Las Vegas, NV 89154, USA}

\author[0000-0003-2251-0602]{John M. Carpenter}
\affiliation{Joint ALMA Observatory, Avenida Alonso de C{\'o}rdova 3107, Vitacura, Santiago, Chile}

\author[0000-0002-1199-9564]{Laura M. P\'erez}
\affiliation{Departamento de Astronom\'ia, Universidad de Chile, Camino El Observatorio 1515, Las Condes, Santiago, Chile}

\author[0000-0002-0786-7307]{Richard Teague}
\affiliation{Center for Astrophysics \textbar\ Harvard \& Smithsonian, 60 Garden Street, Cambridge, MA 02138, USA}

\author[0000-0001-8061-2207]{Andrea Isella}
\affiliation{Department of Physics and Astronomy, Rice University, 6100 Main Street, Houston, TX 77005, USA}

\author[0000-0001-8123-2943]{Luca Ricci}
\affiliation{Department of Physics and Astronomy, California State University Northridge, 18111 Nordhoff Street, Northridge, CA 91130, USA}

\begin{abstract}
We present a detailed analysis for a subset of the high resolution ($\sim$35 mas, or 5 au) ALMA observations from the Disk Substructures at High Angular Resolution Project (DSHARP) to search for faint 1.3 mm continuum emission associated with dusty circumplanetary material located within the narrow annuli of depleted emission (gaps) in circumstellar disks.  This search used the \citet{jennings20} {\tt frank} modeling methodology to mitigate contamination from the local disk emission, and then deployed a suite of injection--recovery experiments to statistically characterize point-like circumplanetary disks in residual images.  While there are a few putative candidates in this sample, they have only marginal local signal-to-noise ratios and would require deeper measurements to confirm.  Associating a 50\%\ recovery fraction with an upper limit, we find these data are sensitive to circumplanetary disks with flux densities $\gtrsim$\,50--70 $\mu$Jy in most cases.  There are a few examples where those limits are inflated ($\gtrsim$\,110 $\mu$Jy) due to lingering non-axisymmetric structures in their host circumstellar disks, most notably for a newly identified faint spiral in the HD 143006 disk.  For standard assumptions, this analysis suggests that these data should be sensitive to circumplanetary disks with dust masses $\gtrsim$\,0.001--0.2 M$_\oplus$.  While those bounds are comparable to some theoretical expectations for young giant planets, we discuss how plausible system properties (e.g., relatively low host planet masses or the efficient radial drift of solids) could require much deeper observations to achieve robust detections. 
\end{abstract}
\keywords{protoplanetary disks --- circumstellar matter --- planets and satellites: formation}

\section{Introduction \label{sec:intro}}

In just the past few years, the observational landscape of planet formation research has expanded dramatically.  New measurements at very high spatial resolution ($\sim$few au) have revealed that protoplanetary disks are richly {\it substructured} (see \citealt{andrews20} for a review).  Observations with the Atacama Large Millimeter/submillimeter Array (ALMA) demonstrate that the (sub-)mm continuum emission from small particles in these disks is frequently distributed in bright rings, separated by darker gaps \citep{long18,dsharp2,vandermarel19,cieza21}.  Hydrodynamics simulations show that interactions between planets and their birth environments can create these kinds of perturbations \citep{kanagawa15,dong15,bae17}.  Given the locations, widths, and depths of these gaps, such simulations suggest that planets with masses $M_p \approx 0.1$--1 M$_{\rm Jup}$ orbiting their (roughly solar-mass) host stars at distances $a_p \approx 10$--100 au may be common at ages of only $\sim$1--3 Myr \citep[e.g.,][]{jin16,clarke18,dsharp7,lodato19}.

This work continues to be refined and expanded upon with more (sub-)mm/cm continuum data, as well as resolved measurements of scattered starlight \citep{avenhaus18,garufi18} and spectral line emission, where the latter reveals substructures in both the intensities \citep{isella16,dsharp8,law21} and kinematics \citep[see][]{dynamics20}.  Collectively, these constraints on the initial architectures of planetary systems can be compared to the properties of the (older) exoplanet population to inform models of planetary migration and add context to analyses of direct imaging \citep[e.g.,][]{bowler16,nielsen19,vigan20} and microlensing surveys \citep[e.g.,][]{gaudi12,penny19}.     

In addition to these more general insights into the evolution of planetary systems, we are also seeing detailed case studies that quantitatively link disk substructures to planetary perturbers.  The directly imaged planets in the cleared disk cavity around the young star PDS 70 are especially exciting examples \citep{keppler18,keppler19,haffert19}.  Using hydrodynamics simulations of the planetary accretion process as a guide \citep[e.g.,][]{ayliffe09,tanigawa12,szulagyi14}, emission models predict that such planets may be easiest to discover in the mid-infrared ($\sim$5--20 $\mu$m), where the dust in a circumplanetary disk (CPD)\footnote{The morphology of the \fix{circumplanetary} material is uncertain \citep{szulagyi16}, but for simplicity we refer to it as a ``disk".} should outshine the planetary photosphere \citep{zhu15b,eisner15,szulagyi19}.  Direct imaging of that emission may be common in the near future (i.e., with {\it JWST} and $\sim$20--40 m ground-based telescopes), and will provide access to the thermal structures of those CPDs.  A helpful complement is available at (sub-)mm/cm wavelengths, where the optical depths are low enough to give some insight on the CPD (dust) masses \citep[e.g.,][]{zhu18,szulagyi18,wang21}.

Deep imaging with ALMA can reach continuum sensitivities comparable to expectations for CPD masses.  While few dedicated (sub-)mm/cm CPD searches have been attempted, those available provide stringent upper limits \citep{isella14,sperez19c,pineda19}.  Recent ALMA observations have identified mm continuum emission from the CPD associated with the PDS 70c planet \citep{isella19,benisty21}.  This kind of information offers unique insights on how much material is available to form planetary satellites, and provides important boundary conditions for understanding the planetary accretion process.

In this article, we describe an attempt to quantify the constraints on any mm continuum emission from CPDs associated with the disk gaps identified by the Disk Substructures at High Angular Resolution Project (DSHARP; \citealt{dsharp1}).  Section \ref{sec:data} presents the selection criteria and the data used in this effort.  Section \ref{sec:disk_removal} develops a methodology to mitigate confusion from local circumstellar disk emission.  Section \ref{sec:CPDs} describes the techniques used to quantify the sensitivity to CPD emission, and Section \ref{sec:results} presents the results.  Finally, Section \ref{sec:discussion} discusses the outcomes of this analysis in the context of simple CPD models and the future prospects for further work.  Section \ref{sec:summary} summarizes the principal results.

\section{Data} \label{sec:data}

\subsection{Selection Criteria}
The full DSHARP sample includes 20 targets observed with ALMA at 240 GHz (1.25 mm) to a roughly uniform depth and angular resolution \citep{dsharp1}.  Our focus in this article was a search for CPDs in the primarily symmetric examples of annular gaps in the mm continuum emission distributions of these disks.  Therefore, we excluded targets where the emission has a dominant non-axisymmetric morphology (i.e., the cases with global spiral patterns; see \citealt{dsharp3,dsharp4}).  We further limited the sample to include only gaps that are at least marginally resolved over the full azimuthal range \citep{dsharp2}.  This latter criterion excluded cases with high inclination angles (e.g., the DoAr 25 or HD 142666 disks), presumably because their emission surfaces are vertically elevated.    

These selection priorities were entirely practical, designed only to facilitate the CPD search methodology discussed below.  They do not imply that excluded cases are any less likely to host CPDs.  The resulting sample includes 9 disk targets, but focuses the CPD searches on 13 individual gaps.  Table \ref{table:data} lists some basic properties of the data associated with the search.    

\begin{deluxetable}{l | c c | c c c}[t!]
\tablecaption{Sample and Data Summary \label{table:data}}
\tablehead{
\colhead{Disk} &
\colhead{$d$} & 
\colhead{Gaps} & 
\colhead{$\theta_{\rm bm}$} &
\colhead{PA$_{\rm bm}$} &
\colhead{RMS} \\
\colhead{} & 
\colhead{(pc)} & 
\colhead{} & 
\colhead{(mas)} &
\colhead{(\degr)} & 
\colhead{($\frac{\mu{\rm Jy}}{{\rm bm}}$)} \\
\colhead{(1)} & \colhead{(2)} & \colhead{(3)} & \colhead{(4)} & \colhead{(5)} & \colhead{(6)}
}
\startdata
SR 4      & 135 & D11       & $35{\times}34$ & \phn\phn5 & 18 \\
RU Lup    & 158 & D29       & $26{\times}25$ & 145       & 14 \\
Elias 20  & 138 & D25       & $31{\times}23$ & \phn76    & 10 \\
Sz 129    & 160 & D41, D64  & $44{\times}31$ & \phn87    & 12 \\
HD 143006 & 167 & D22, D51  & $48{\times}45$ & \phn51    & 10 \\
GW Lup    & 155 & D74       & $45{\times}43$ & \phn\phn1 & 10 \\
Elias 24  & 139 & D57       & $37{\times}34$ & \phn88    & 12 \\
HD 163296 & 101 & D48, D86  & $48{\times}38$ & \phn79    & 19 \\
AS 209    & 121 & D61, D97$^\dagger$ & $38{\times}37$ & 106       & 10 \\
\enddata
\tablecomments{(1) Target name; (2) distance based on the {\it Gaia} EDR3 parallax \citep{gaia,gaia_edr3}; (3) \citet{dsharp2} gap designation (for reference, `D' refers to a `dark' feature, and the associated number corresponds to the radius of the gap center in au); (4) and (5) FWHM dimensions and position angle of the synthesized ALMA beam; and (6) RMS noise in the synthesized image, measured in an annulus 1.2 $R_{\rm out}$ (see Table \ref{table:geom} and associated discussion) to 4\farcs25 from disk center. \\
$^\dagger$ Following the interpretations of \citet{dsharp8} and \citet{dsharp7}, we chose to designate the wide outer gap in the AS 209 disk as `D97', effectively combining the D90 and D105 gap designations of \citet{dsharp2}.} 
\end{deluxetable}

\subsection{Data Processing}
For each disk, we retrieved the publicly available self-calibrated (pseudo)continuum visibilities from the measurement sets in the DSHARP data repository\footnote{\url{https://almascience.org/almadata/lp/DSHARP}} (see \citealt{dsharp1} for calibration details).  Those visibilities were spectrally averaged to one channel per spectral window and time-averaged into 30 s intervals to reduce the data volume.  This averaging is modest enough to avoid bandwidth- or time-smearing effects.  All of the post-processing of these data was conducted with the {\tt CASA} v5.7 package  \citep{mcmullin07}.  

Throughout the analysis described below, imaging associated with these visibility data was performed following the DSHARP script recommendations, but with three minor modifications.  First, we used 2$\times$ larger pixel sizes (6 mas, or $\sim$0.2$\times$ the typical FWHM of the point-spread function, or PSF).  Second, we adopted deeper {\sc clean} thresholds, equivalent to twice the RMS values listed in Table \ref{table:data}.  And third, a final step to the imaging process was included to treat an intrinsic deficiency in the {\tt CASA/tclean} algorithms associated with {\sc clean} beam restoration for data with complicated PSFs due to the combination of disparate antenna configurations.  The issue and its solution are described by \citet{jorsater95} and \citet{czekala21}.  To summarize, an additional correction step rescales the {\sc clean} residuals by the ratio $\epsilon$ of the Gaussian {\sc clean} beam to the PSF before adding them to the convolved {\sc clean} model, to ensure that the appropriate units (Jy per {\sc clean} beam) are propagated.  We found a typical rescaling factor $\epsilon \approx 0.65$ (with extremes of 0.4 and 0.8).  The updated RMS values are listed in Table \ref{table:data}, measured as described by \citet{dsharp1}.

\section{Mitigating Contamination from Circumstellar Disk Emission} \label{sec:disk_removal}

Contrast with the local circumstellar disk emission is the most formidable challenge for identifying faint CPDs embedded in narrow gaps.  The ambient material emits a comparatively bright continuum, and PSF convolution smears that emission into the gap and complicates the CPD search.  Moreover, {\sc clean} artifacts produced by the sparse distributions of ALMA antennas with long baselines might mimic CPD emission (e.g., manifested as ``spokes" in gaps; see \citealt{andrews16,dsharp1}).  Ideally, these risks can be mitigated by searching for CPDs in {\it residual} products, where the emission from the circumstellar disk has been removed.  The following sections describe our approach to this task and its outcomes.

\subsection{Modeling Procedure}
In this specific context, the detailed morphology of the circumstellar disk emission is irrelevant so long as there is some model available that does a good job of accounting for (and thereby removing) it.  This means a sophisticated statistical inference of model parameters is not desirable.  That is helpful, since developing an appropriate parameterization and comparing it to the data can be technically challenging and incredibly time-consuming.  Instead, we adopted the approach introduced by \citet{jennings20,jennings21} and implemented in the associated {\tt frank} software package.  Assuming an intrinsically one-dimensional (radial) emission distribution, {\tt frank} uses a Gaussian Process to rapidly optimize a model of the (deprojected, real) visibilities.  The fundamental assumption is that this axisymmetric modeling excludes CPD emission by default: because it should be both faint and azimuthally localized, the one-dimensional nature of the model averages out any CPD contribution at that radius.  We confirmed a posteriori that this is indeed the case (see Section \ref{sec:CPDs}).

To construct such a model, we first explored the effects of the {\tt frank} hyper-parameters, fixing the disk geometry parameters (see below) to the values measured by \citet{dsharp2}.  The $R_{\rm max}$ hyper-parameter (beyond which {\tt frank} assumes there is zero emission) was set to twice the radius of the outer extent of the observed emission, $R_{\rm out}$, measured in the synthesized image and defined by the contour that reaches twice the RMS noise (see Table \ref{table:data}).  The $N$ and $p_0$ hyper-parameters, which define the radial gridding and regularize the emission power spectrum, were kept at their default values (300 and 10$^{-15}$ Jy$^2$; see \citealt{jennings20}).  Combinations over a range of the hyper-parameters $\alpha$ and $w_{\rm smooth}$, which act like prior weights to regularize variations in the model brightness distribution, were explored interactively.  We favored combinations that damp high frequency oscillations in the brightness profile, and set $\alpha{=}1.3$ and $w_{\rm smooth}{=}0.1$ for all disks except AS 209, where instead $w_{\rm smooth}{=}0.001$.  In any case, the outcomes for this study are not affected by these selections.

Having set the hyper-parameters, we then refined estimates for the four geometric parameters that describe the emission morphology projection into the sky-plane: inclination ($i$, where $i = 0\degr$ for face-on and 90\degr\ for edge-on), position angle of the major axis (PA, measured E of N), and the right ascension and declination offsets from the observed phase center ($\Delta x$ and $\Delta y$, where positive values are to the E and N, respectively).  We relied on a visual selection of these parameters, examining the images synthesized from the residual visibilities for models that spanned small grids around the \citet{dsharp2} estimates (using 0.5 mas and 0.5\degr\ steps for the offsets and projection angles, respectively).  Unfortunately, automatic optimization approaches to this task performed poorly in the general case of lingering asymmetries, but such obstacles can be overcome with some (admittedly tedious) experience.  Some insights on the relevant issues are shared in Appendix \ref{app:resid}.  Table \ref{table:geom} lists the adopted geometric parameters.  

\begin{deluxetable}{l | c c c c | c}[t!]
\tablecaption{Circumstellar Disk Emission Geometries \label{table:geom}}
\tablehead{
\colhead{Disk} &
\colhead{$\Delta x$} & 
\colhead{$\Delta y$} &
\colhead{$i$} & 
\colhead{PA} & 
\colhead{$R_{\rm out}$} \\
\colhead{} & 
\colhead{(mas)} & 
\colhead{(mas)} & 
\colhead{(\degr)} & 
\colhead{(\degr)} &
\colhead{(\arcsec)} \\
\colhead{(1)} & \colhead{(2)} & \colhead{(3)} & \colhead{(4)} & \colhead{(5)} & \colhead{(6)}
}
\startdata
SR 4      & $-60$  & $-509$ & 22 & \phn18\phn & \phn0.25 \\
RU Lup    & $-17$  & $+86$  & 19 & 121\phn    & \phn0.42 \\
Elias 20  & $-52$  & $-490$ & 54 & 153\phn    & \phn0.48 \\
Sz 129    & $+5$   & $+6$   & 32 & 153\phn    & \phn0.48 \\
HD 143006 & $-6$   & $+23$  & 16 & 167\phn    & \phn0.53 \\
GW Lup    & $+0.5$ & $-0.5$ & 39 & \phn37\phn & \phn0.75 \\
Elias 24  & $+107$ & $-383$ & 30 & \phn45\phn & \phn1.05 \\
HD 163296 & $-3.5$ & $+4.0$ & 47 & 133\phn    & \phn1.08 \\
AS 209    & $+2$   & $-3$   & 35 & \phn86\phn & \phn1.20 \\
\enddata
\tablecomments{(1) Target name; (2) and (3) RA and declination offsets from the phase center; (4) inclination angle; (5) position angle; (6) outer (radial) boundary of the observed emission (the deprojected contour radius for twice the RMS noise in Table \ref{table:data}).  Formal uncertainties on the geometric parameters were not measured, but approximate estimates based on the residual map appearances suggest a precision of $\sim$1--2 mas in the offsets and $\sim$1\degr\ in the projection angles (up to perhaps $\sim$2--3\degr\ for the more face-on geometries).  The estimated $R_{\rm out}$ values have a $\sim$10\%\ uncertainty.}
\end{deluxetable}

\begin{figure}[t!]
\includegraphics[width=\linewidth]{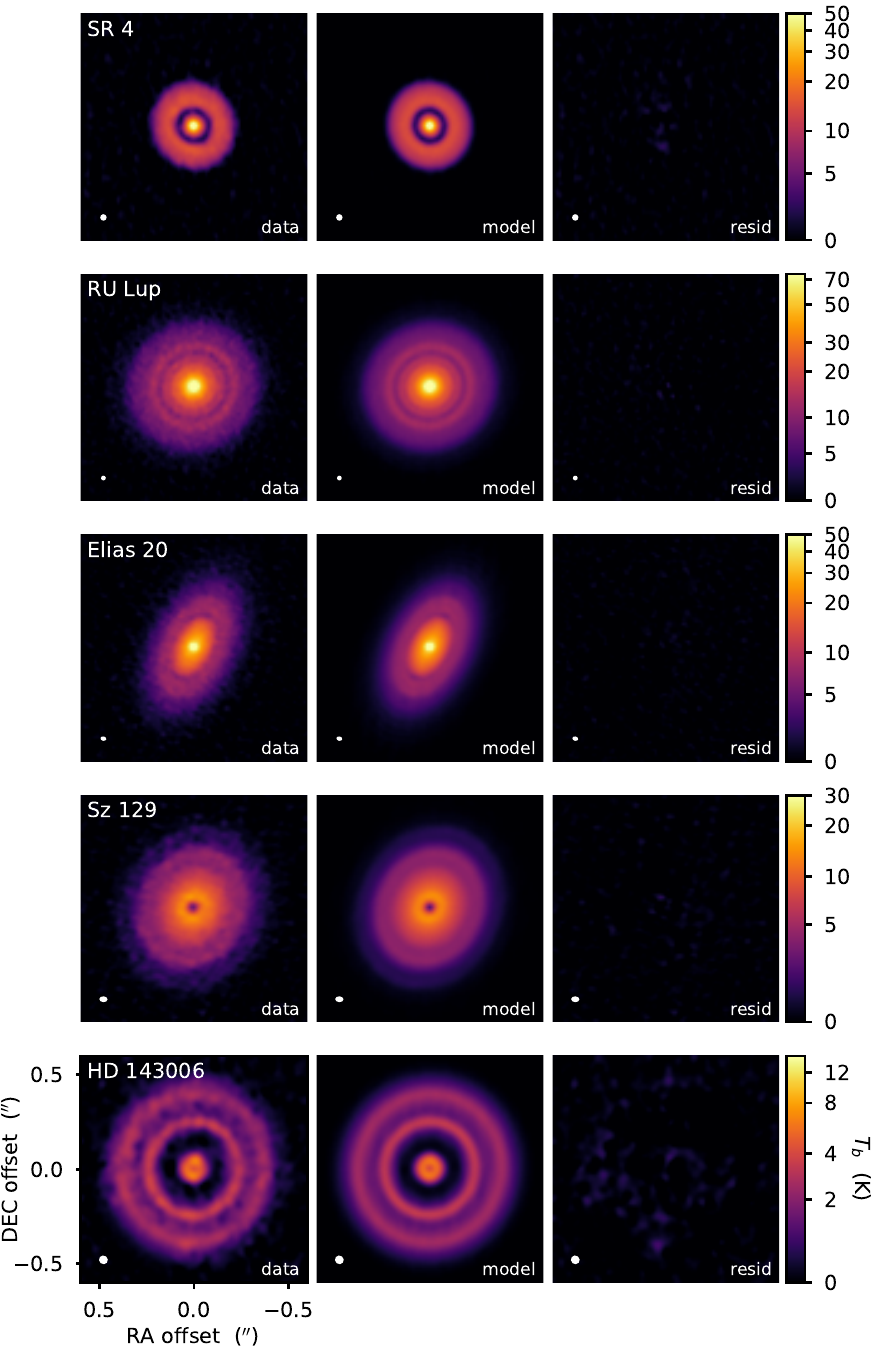}
\caption{Comparisons of the data, models, and residuals for the smaller circumstellar disks (see Figure \ref{fig:dmrs2} for the larger disks) in the image plane.  The model and residual images were constructed from the {\tt frank} model and residual visibilities in the same way as for the data.  The {\sc clean} beam dimensions are marked in the lower left corner of each panel. The data, model, and residual images of a given disk are shown on the same brightness temperature scale (with an asinh stretch), assuming the Rayleigh-Jeans approximation.  The HD 143006 images were constructed from the {\it revised} visibilities (see the text and Appendix \ref{app:asym}).  Figure \ref{fig:resid_maps} shows a more detailed examination of the residual images.      
\label{fig:dmrs1}
}
\end{figure}

\begin{figure}[t!]
\includegraphics[width=\linewidth]{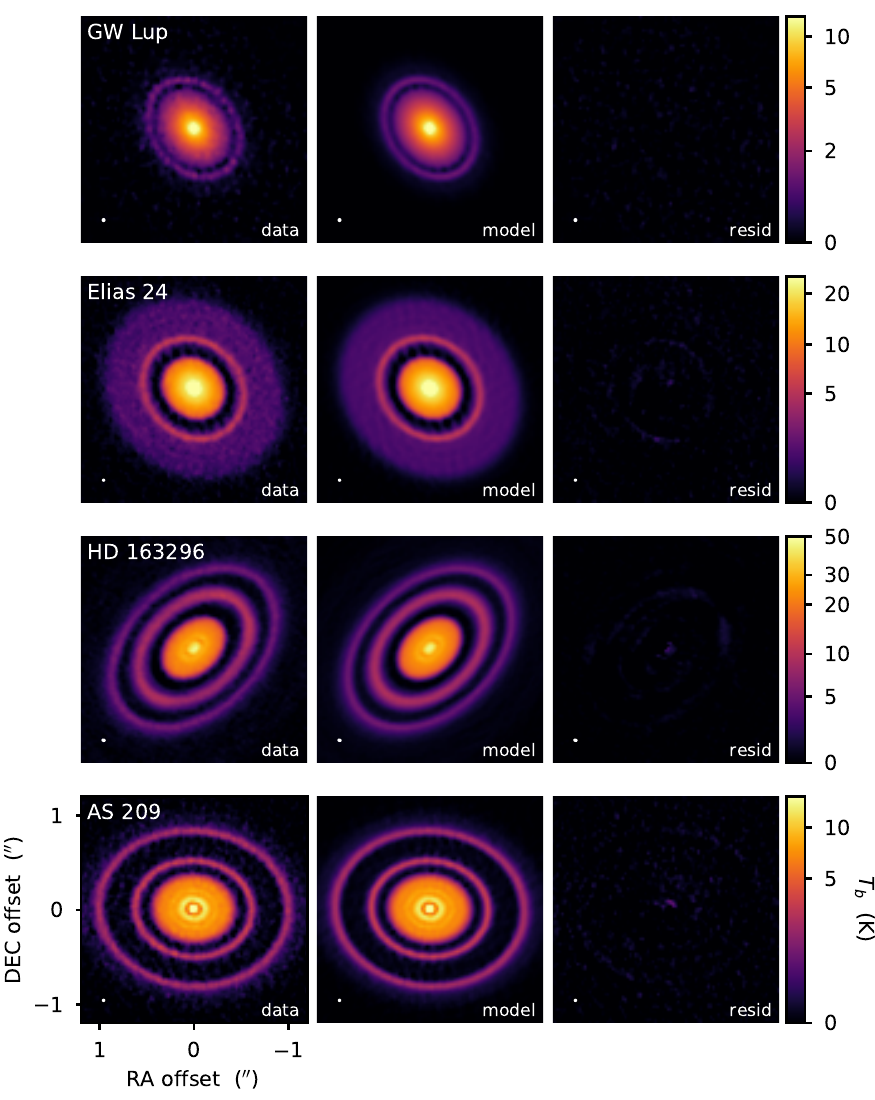}
\caption{Additional image-plane comparisons of the data, models, and residuals for the larger disks in this sample, as in Figure \ref{fig:dmrs1} (note the 2$\times$ larger spatial extent in each panel).  The HD 163296 images were constructed from the {\it revised} visibilities (see the text and Appendix \ref{app:asym}).
\label{fig:dmrs2}
}
\end{figure}

Finally, the visibility data were modeled using {\tt frank} with the adopted hyper-parameters and disk geometry parameters fixed.  For two of the nine targets, the disks around Elias 20 and GW Lup, we followed the ``point source correction" procedure outlined by \citet{jennings21}.  This involved removing central point-like contributions of 0.61 and 0.48 mJy (as measured for deprojected baselines longer than 5.7 and 7.0 M$\lambda$), respectively, performing the {\tt frank} modeling, and then adding those contributions back into the best-fit models.  These corrections improved the quality of the model brightness profiles by damping oscillatory artifacts, but since we are focused on the widest gaps, they had no real effect on the interpretations outlined in Section \ref{sec:CPDs}.  For two other targets, the disks around HD 163296 and HD 143006, we needed an additional step to account for their bright, azimuthally-localized asymmetries \citep{dsharp9,dsharp10}.  The emission from these features was first removed based on an excised portion of the {\sc clean} model, following the procedure described in Appendix \ref{app:asym}.  The results were {\it revised} visibility datasets for ``symmetric" emission distributions, which were then modeled with {\tt frank} as outlined above.

\subsection{Modeling Results} \label{sec:modeling_results}

\begin{figure*}[t!]
\includegraphics[width=\linewidth]{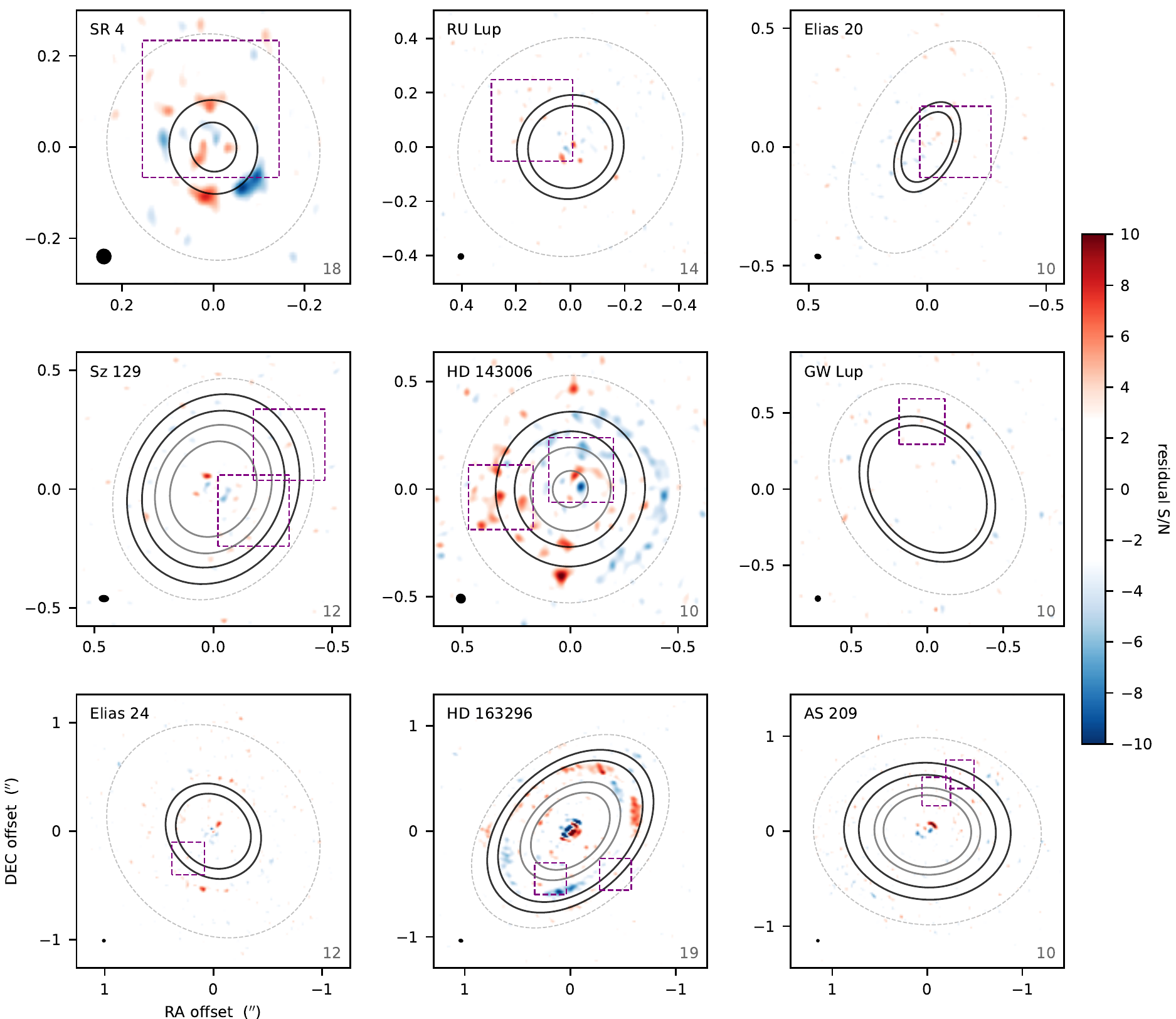}
\caption{A more detailed view of the images synthesized from the residual visibilities.  These are the same images as in the rightmost panels in Figures \ref{fig:dmrs1} and \ref{fig:dmrs2}, but shown here together with the same linear stretch on a diverging color-scale for each target.  The scale represents the residual S/N, where each map has been normalized by the RMS noise determined in an annulus spanning 1.2 $R_{\rm out}$ to 4\farcs25 from the disk center (with values marked in the lower right corner of each panel in $\mu$Jy bm$^{-1}$ units; see Table \ref{table:data}).  These representations better illustrate the low-level asymmetries that are not accounted for in the {\tt frank} modeling.  Solid black ellipses (and gray, for the four cases with multiple gaps) mark the gap ``boundaries", defined as described in the text.  The dashed gray ellipse denotes the outer boundary of the mm continuum emission ($R_{\rm out}$ in Table \ref{table:geom}).  The purple dashed boxes mark regions 0\farcs3 on a side centered on the peak residual in each gap; a gallery of zoomed-in views of these boxes is shown in Figure \ref{fig:peaks}.  The synthesized beam dimensions are shown as filled black ellipses in the lower left corner of each panel.
\label{fig:resid_maps}
}
\end{figure*}

\begin{deluxetable}{l l | c c c | c c}[t!]
\tablecaption{Estimated Gap Properties \label{table:gaps}}
\tablehead{
\colhead{Disk} & 
\colhead{Gap} &
\colhead{$r_{\rm gap}$} & 
\colhead{$\sigma_{\rm gap}$} &
\colhead{$\delta_{\rm gap}$} & 
\colhead{$T_0$} & 
\colhead{$q$} \\
\colhead{} & 
\colhead{} & 
\colhead{(mas)} &
\colhead{(mas)} &
\colhead{} & 
\colhead{(K)} & 
\colhead{} \\
\colhead{(1)} & \colhead{(2)} & \colhead{(3)} & \colhead{(4)} & \colhead{(5)} & \colhead{(6)} & \colhead{(7)} 
}
\startdata
SR 4      & D11  & \phn79 & 10  & 30  & \phn17    & 0.70 \\
RU Lup    & D29  & 179    & 10  & 2   & \phn25    & 0.80 \\
Elias 20  & D25  & 181    & 11  & 1.8 & \phn21    & 1.25 \\
Sz 129    & D41  & 243    & 20  & 1.3 & \phn12    & 0.90 \\
Sz 129    & D64  & 376    & 20  & 1.8 & 170       & 3.30 \\
HD 143006 & D22  & 140    & 35  & 20  & \phn\phn7 & 0.50 \\
HD 143006 & D51  & 315    & 26  & 5   & \phn\phn7 & 0.50 \\
GW Lup    & D74  & 485    & 14  & 10  & \phn10    & 1.25 \\
Elias 24  & D57  & 420    & 35  & 40  & \phn23    & 0.75 \\
HD 163296 & D48  & 490    & 40  & 50  & \phn39    & 0.70 \\
HD 163296 & D86  & 850    & 41  & 12  & \phn48    & 0.90 \\
AS 209    & D61  & 510    & 32  & 25  & \phn15    & 0.70 \\
AS 209    & D97  & 800    & 62  & 30  & \phn15    & 0.70 \\
\enddata
\tablecomments{(1) Target name; (2) gap designation from \citet{dsharp2}; (3) gap center; (4) gap width (Gaussian standard deviation); (5) gap depth (multiplicative depletion factor); (6) normalization of the local background power-law profile (brightness temperature at 0\farcs1); (7) background profile power-law index.  See Appendix \ref{app:SBprof} for the model and parameter definitions.
}
\end{deluxetable}

This approach generally performed well.  To illustrate the quality of the disk emission removal, Figures \ref{fig:dmrs1} and \ref{fig:dmrs2} compare the images synthesized from the data, model, and residual visibilities for each disk on the same spatial and intensity scales.  A direct comparison of the data and model visibilities, along with the inferred model brightness profiles, is presented in Appendix \ref{app:SBprof}.  

We used the model brightness profiles derived with {\tt frank} to estimate the locations, widths, and depths of the gaps of interest, presuming a Gaussian morphology superposed on a background power-law profile.  The parameters of this model -- particularly the means ($r_{\rm gap}$), standard deviations ($\sigma_{\rm gap}$), and depletion factors (i.e., amplitudes; $\delta_{\rm gap}$) of those Gaussian gaps -- are catalogued in Table \ref{table:gaps} (see Appendix \ref{app:SBprof} for details).  These are meant only as rough guidance: many of the gaps are not Gaussian and the backgrounds are not all described well with power-laws.  The gap centers are accurate, but the widths and depths are crude estimates.   

Close inspections of the residual images reveal some lingering structure that was imperfectly captured by the {\tt frank} modeling.  Figure \ref{fig:resid_maps} shows a more instructive view of the images synthesized from the residual visibilities to illustrate those features.  These images are on a common S/N scale, normalized by the map RMS listed in Table \ref{table:data}, and are annotated to reference the locations of the gaps of interest and $R_{\rm out}$.  The gap boundaries are intended to illustrate what is seen in the (beam-convolved) data images; they correspond to $r_{\rm gap} \pm (\sigma_{\rm gap} + \sigma_{\rm bm})$, where $\sigma_{\rm bm} = \langle \theta_{\rm bm} \rangle / 2\sqrt{2 \ln{2}}$ represents the (geometric mean) width of the synthesized beam (see Table \ref{table:data}). 

Figure \ref{fig:resid_maps} demonstrates that the model residuals generally have low S/N (white is $|{\rm S/N}| \le 3$ in these images), particularly within the gap regions of interest here.  Most cases show lingering residuals at $\sim$4--7$\times$ the RMS noise near the disk centers (within $\sim$0\farcs1).  In some sense, this is expected for models with limited fidelity.  We found that the {\it maximum} imaged residuals from this methodology were typically only ($\pm$) a few percent of the observed emission levels in the data inside 0\farcs1 (increasing to $\sim$10\%\ in the fainter outer regions).  However, the emission peaks in these regions have ${\rm S/N} \approx 200$--400, so a few percent deviation will produce a residual S/N of order $\sim$10.  These inner disk features might be caused by radiative transfer effects (i.e., a vertically flared emitting surface; see below and Appendix \ref{app:resid}) or genuine asymmetries, but in most cases it would require deeper observations (ideally at higher resolution) to know for sure.

\subsection{Notable Residuals}

The residual maps for three cases merit specific mention (see also \citealt{jennings21}).  The SR 4 disk exhibits some of the strongest residuals ($\sim$8$\times$ the RMS noise).  This is the most compact disk in the sample, with the innermost gap of interest here.  The scale of these features suggest they are analogues of the inner disk residuals noted above for other targets, although it is interesting that they primarily reside at the outer edge of the gap.  Perhaps that is a clue that the residuals generally found close to the disk centers are associated with non-axisymmetric substructures that are not easily recognized because they suffer from poor resolution and/or PSF convolution artifacts.  The HD 163296 disk also exhibits strong and structured residuals, both in the inner $\sim$0\farcs1 and associated with the bright ring (B67) separating the D48 and D86 gaps.  The morphological pattern of these residuals suggests a ring surface that is inclined and/or elevated with respect to the global mean values (e.g., Appendix \ref{app:resid}; see also \citealt{dsharp9}).  \citet{doi21} reached similar conclusions from a detailed modeling analysis of the same data.  A comparable, albeit much weaker, pattern is visible well outside the D57 gap for the Elias 24 disk, perhaps associated with an analogous deviation in the projected emission surface height for the B77 bright ring.

\begin{figure}[t!]
\includegraphics[width=\linewidth]{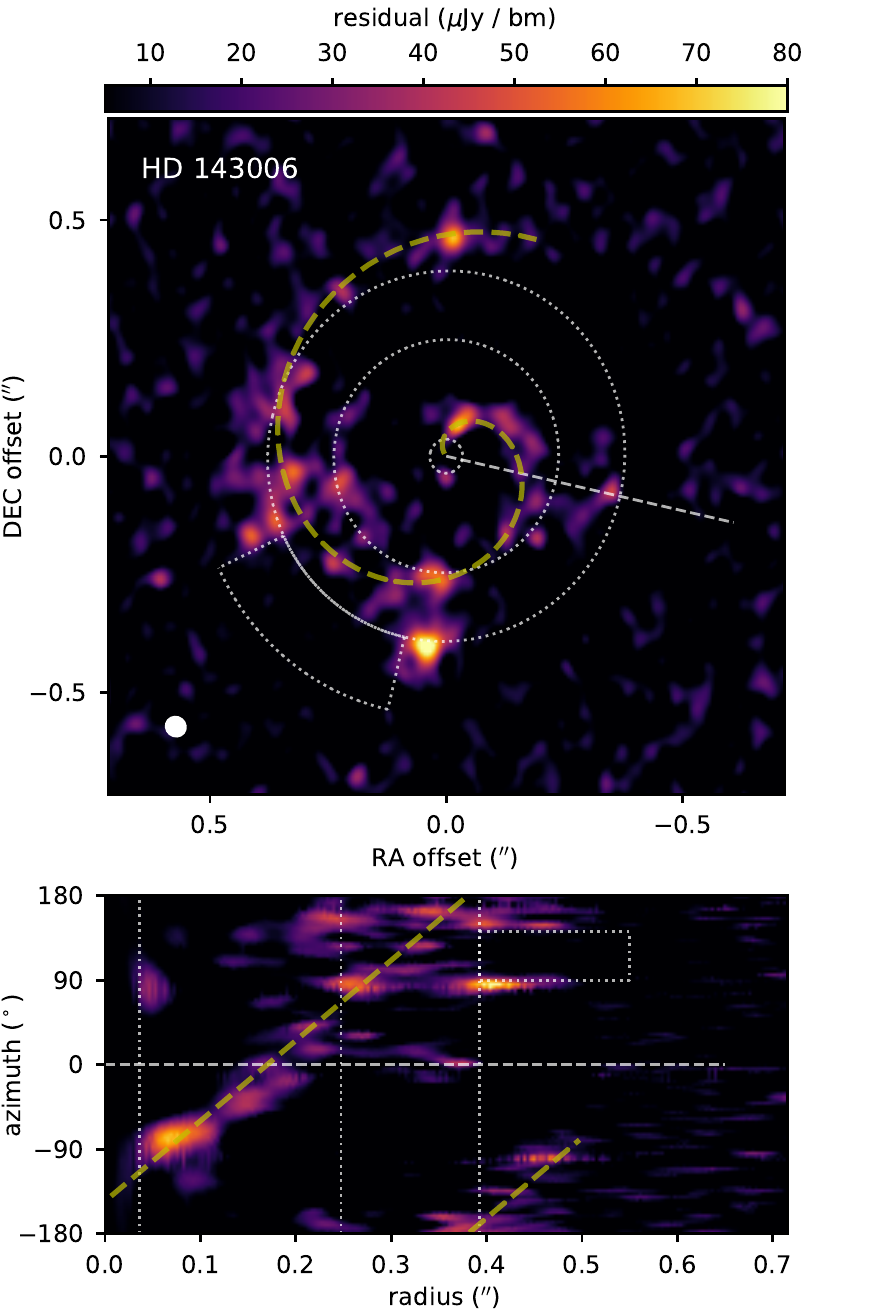}
\caption{The residual image (top) and its deprojection into disk-frame polar coordinates (bottom), following the formalism of \citet{dsharp2}, on a linear intensity scale to highlight the low-level spiral.  The bright ring emission peaks and the region around the pronounced azimuthal asymmetry are marked in the sky-plane (polar deprojected) image with white ellipses (vertical lines) and an arc (rectangle), respectively.  For reference, the straight dashed lines in each panel mark zero azimuth ($\varphi = 0$\degr; with increasing $\varphi$ clockwise in the sky plane).  The spiral model described in the text is overlaid as a dashed yellow curve, to guide the eye.
\label{fig:HD143_spiral}
}
\end{figure}

The residuals for the HD 143006 disk are especially complex in Figure \ref{fig:resid_maps}.  A close examination of the residual map shows a noisy, one-armed spiral pattern (the $+/-$ residual patterns in Figure \ref{fig:resid_maps} are artifacts of the axisymmetric {\tt frank} modeling: the negative [blue] residuals are the mirror image of the spiral feature).  This is perhaps illustrated more clearly in Figure \ref{fig:HD143_spiral}, where the polar deprojection (in the disk-plane polar coordinate system) of the image is also included.  A (visually tuned, not fitted) Archimedean spiral\footnote{We were unable to find a logarithmic spiral model that performed well across the full azimuthal range of the feature.} morphology approximates this residual feature fairly well, with $r_{\rm spiral} \approx 0.170 + 0.067 \varphi$ (in units of arcseconds), where the azimuth $\varphi$ (in radians) follows the convention outlined by \citet{dsharp2}.  This spiral apparently traverses through all of the rings and gaps in the disk, spanning $\sim$420\degr\ in azimuth; it is accompanied by lingering clumpy residuals that are spatially coincident with the southeastern quadrants of the B41 and B65 rings.  \citet{dsharp10} noted the innermost portion of the feature as a ``bridge" between the B8 and B41 rings (at PA$\approx$300\degr).  Their residual map has a more muted peak at the innermost extension of the arm, in part because they modeled the B8 ring with a different $i$ (i.e., a warp).  For the orientation favored by interpretations of the scattered light shadows \citep{benisty18} and gas kinematics \citep{dsharp10}, with the east side of the disk nearer to the observer, the spiral is trailing.  Given the low S/N of this spiral pattern, it is unclear what to make of this feature.  We offer some brief speculation in Section \ref{sec:discussion}, but deeper observations that can facilitate a more quantitative analysis are needed before drawing any conclusions.

\section{Assessing Sensitivity to CPD Emission} \label{sec:CPDs}

After the bright (and presumed symmetric) emission from the circumstellar disk was modeled and removed, the residual images presented in Figure \ref{fig:resid_maps} could be used to search for faint CPD emission in the gaps of interest.  In our analysis, we assumed that the regions most likely to exhibit CPD emission were the annular zones bounded by $r_{\rm gap}{\pm}\sigma_{\rm gap}$ in a disk-frame coordinate system.  Moreover, we implicitly assumed that the optical depths in these gap regions were low enough that they did not obscure any potential CPD emission (which would then be missing after the {\tt frank} model subtraction).  This section presents the methodology adopted for a statistical assessment of our sensitivity to CPD emission in those search zones.  An analysis of the outcomes from those assessments and evaluations of the residual peaks are presented in Section \ref{sec:results}.

\begin{figure*}[t!]
\includegraphics[width=\linewidth]{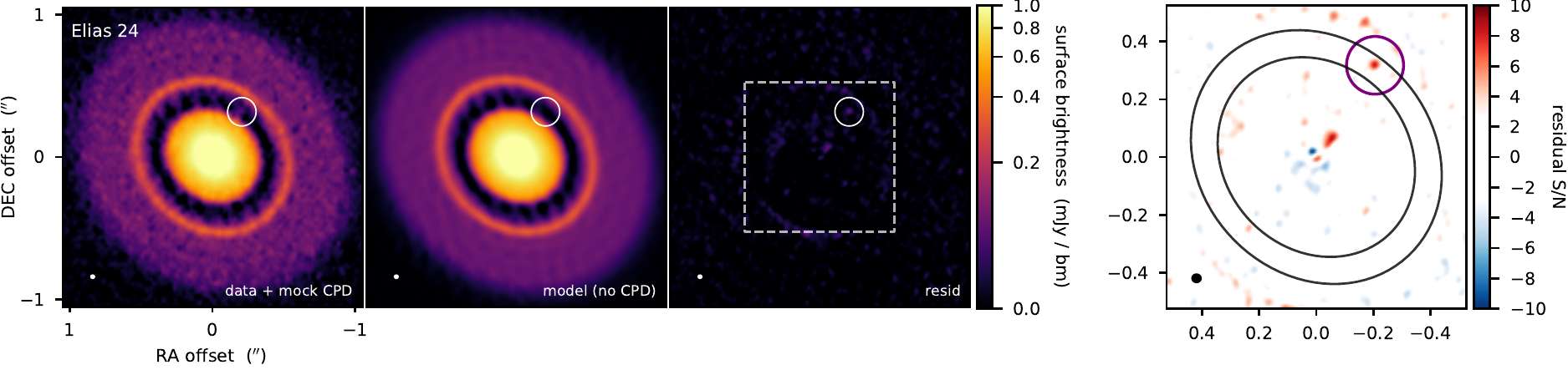}
\caption{An image-plane illustration of one iteration in the mock CPD injection--recovery analysis for the Elias 24 D57 gap.  The three panels on the left show the data with an injected CPD signal, the axisymmetric {\tt frank} model, and the corresponding residuals, as in Figure \ref{fig:dmrs2} (but with scale and stretch adjusted for clarity).  The rightmost panel shows the residual S/N map, as in Figure \ref{fig:resid_maps}.  The search annulus corresponding to $r_{\rm gap}{\pm}\sigma_{\rm gap}$ is marked with black ellipses.  This image has a more compact spatial range, as marked by the dashed square in the third panel from left.  This residual S/N map more clearly highlights the $F_{\rm cpd} \approx 100$\,$\mu$Jy mock CPD to the northwest (circled in purple here, and white in the other panels).  Note the low-level ``spoke" artifacts crossing the gap in both the data and model panels, affirming the concerns that motivated the modeling efforts in Section \ref{sec:disk_removal}.  The mock CPD cannot be recovered in the original data image without that modeling and removal procedure. 
\label{fig:EL24_demo}
}
\end{figure*}

A simplistic upper limit based only on the RMS residual scatter in a gap annulus is generally an inappropriate estimator for the sensitivity to CPD emission, since the distribution of pixel intensities is often too correlated (i.e., the gap annuli are not covered by a sufficient number of independent resolution elements to use standard Gaussian statistics) and can be significantly skewed by lingering non-axisymmetric residuals.  Instead, we opted for a statistical approach that assessed the ability to faithfully recover injected (mock) CPD signals.   

For this purpose, we assumed a point source model for the mock CPD emission.  Theoretical models suggest that CPDs are truncated at radii of $\sim$0.1--0.5~$R_{\rm H}$, where the Hill radius $R_{\rm H} \approx a_p \, (M_p / 3 M_\ast)^{1/3}$ \citep[e.g.,][]{quillen98,martin11}.  Presuming giant planets ($M_p \approx {\rm M}_{\rm Jup}$) near the gap centers ($a_p \approx r_{\rm gap}$), any CPDs in this sample are indeed expected to have diameters smaller than the ALMA resolution.  A mock CPD model has three parameters: a location ($r_{\rm cpd}$, $\varphi_{\rm cpd}$; polar coordinates in the disk-frame) and a flux density ($F_{\rm cpd}$).  The mock CPD visibilities for a given set of parameters were computed from the Fourier transform of an offset point source, as described in Appendix \ref{app:CPD_model}. 

Each iteration of the statistical assessment framework we followed has five basic steps:
\begin{enumerate}
    \item Assign mock CPD parameters.  $F_{\rm cpd}$ was selected from a grid spanning 10--250 $\mu$Jy at 10 $\mu$Jy intervals.  Locations were randomly drawn from (independent) uniform distributions such that $r_{\rm cpd} \in (r_{\rm gap} \pm \sigma_{\rm gap})$ and $\varphi_{\rm cpd} \in (\pm180\degr)$. 
    \item Compute mock CPD visibilities (see Appendix \ref{app:CPD_model}) and add them to the {\it observed} ALMA visibilities.
    \item Model these composite data and derive a set of residual visibilities as described in Section \ref{sec:disk_removal}.
    \item Image the residual visibilities (see Section \ref{sec:data}).
    \item Measure the peak in the search zone of the residual image and compare it to the mock CPD inputs.
\end{enumerate}
To build up the mock catalog for each gap, these steps were iterated 500$\times$ per $F_{\rm cpd}$ value.  Figure \ref{fig:EL24_demo} illustrates a typical outcome for this procedure.

While step (1) is self-explanatory, and steps (2) and (3) were described above, the details of steps (4) and (5) merit some discussion.  Readers may (reasonably) wonder why we analyzed residuals in the image plane, rather than performing the search in the Fourier domain.  Indeed, testing the recovery process for mock CPD injections onto a pure Gaussian noise distribution found that the latter option performed equally well, with a reduced computational cost.  However, a visibility-based forward modeling approach to the recovery became problematic in reality, where asymmetric residuals severely bias the outcomes (even at modest S/N).  Attempts to guide the search to ignore those residuals (with increasingly sophisticated priors) ended up becoming complicated efforts to ``model the noise", which encouraged us to instead opt for the simpler image-based alternative.

\begin{figure*}[t!]
\includegraphics[width=\linewidth]{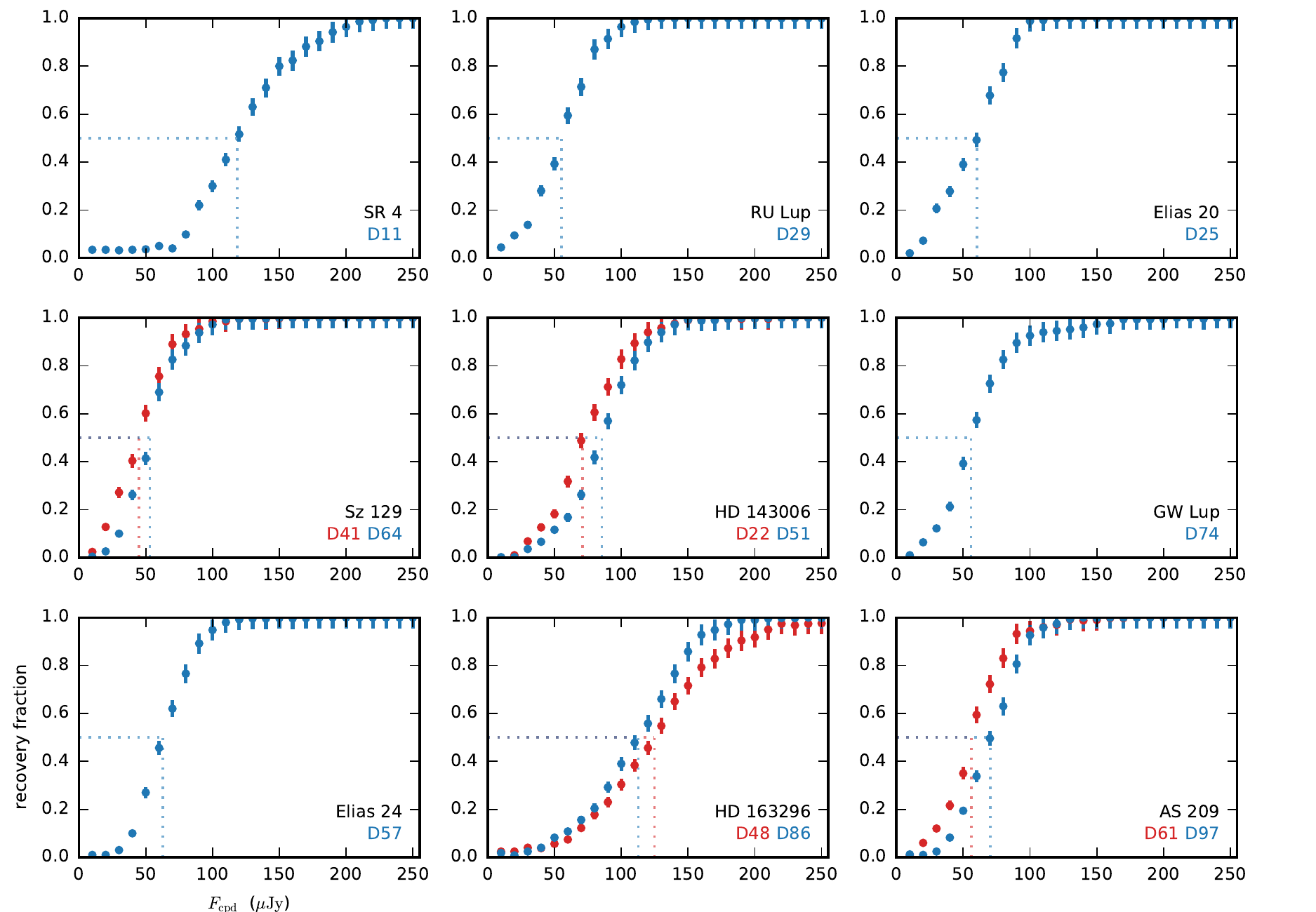}
\caption{The fraction of injected mock CPDs that are successfully recovered as a function of their flux densities.  Each point corresponds to 500 mock injections; error bars reflect only the Poisson uncertainties.  In disks with multiple gaps under consideration, the recovery profiles for the innermost gaps are red (as labeled in the lower right corner of each panel).  The 50\%\ recovery fractions and their corresponding flux densities are marked with dotted lines for each gap of interest.
\label{fig:recovs}
}
\end{figure*}

Ideally, forward modeling the visibilities has the advantage of a well-defined metric for a successful recovery (in terms of a posterior probability).  But in the image-based case, ``success" is subjective.  We used an astrometric criterion, requiring that the peak residual lies within a (sky-projected) distance $\Delta R$ of the injected CPD location.  The adopted distance tolerance was
\begin{equation}
    \begin{aligned}
        \Delta R &= {\rm max}\left( 2 \sigma_{\rm ast}, 12 \, {\rm mas}\right), \,\,\,\,\, {\rm where} \\
        \sigma_{\rm ast} &= 70\arcsec \, \left[ \left(\frac{\nu}{{\rm GHz}}\right) \left(\frac{B_{\rm max}}{{\rm km}}\right) \left(\frac{I_{\rm peak}}{{\rm RMS}_{\rm gap}}\right)\right]^{-1}.
    \end{aligned}
    \label{eq:crit}
\end{equation}
Here, $\sigma_{\rm ast}$ is the formal astrometric uncertainty\footnote{The definition here is based on the \href{https://almascience.nrao.edu/documents-and-tools/cycle7/alma-technical-handbook/view}{ALMA Technical Handbook}, although alternatives  \citep[e.g.,][]{reid88} give similar results.}, 12 mas corresponds to 2 image pixels, $\nu$ is the frequency, $B_{\rm max}$ is the longest baseline, and $I_{\rm peak}/{\rm RMS}_{\rm gap}$ is the recovered peak S/N (see below for details).  

The residual imaging aspect of the recovery procedure (step 4 above) was especially time-consuming.  It amounted to running the {\tt CASA/tclean} algorithm some 12,500 times for each gap of interest.  Extensive experimentation demonstrated that some helpful speed gains can be made by imaging with a mask that covered only around the appropriate search annulus ($r_{\rm gap}{\pm}1.5 \sigma_{\rm gap}$), without any negative effects on the recoveries.  This required us to modify the adopted {\tt scales} parameter in each case, starting from `point-like' ({\tt scales=0}) and incrementing by 5 pixel intervals ($\sim$1 FWHM of the PSF) until exceeding the mask annulus width ($3 \, \sigma_{\rm gap}$).

\section{Results} \label{sec:results}

The key results of the mock CPD injection--recovery exercise described in Section \ref{sec:CPDs} are summarized in Figure \ref{fig:recovs}.  These profiles show the fraction of mock CPDs that were recovered (using the criterion described in Equation \ref{eq:crit}) as a function of the injected flux density.

\begin{deluxetable*}{l l | c c c c c c | c c c c c c}[t!]
\tablecaption{Peak Residuals and Mock CPD Recoveries \label{table:resids}}
\tablehead{
\colhead{Disk} & 
\colhead{Gap} &
\colhead{$\Delta \varrho_{\rm peak}$} & 
\colhead{PA$_{\rm peak}$} & 
\colhead{$r_{\rm peak}$} & 
\colhead{$\varphi_{\rm peak}$} &
\colhead{$I_{\rm peak}$} &
\colhead{RMS$_{\rm gap}$} & 
\multicolumn{6}{c}{$F_{\rm cpd}$ ($\mu$Jy) for a given recovery fraction} \\
\colhead{} & 
\colhead{} & 
\colhead{(mas)} & 
\colhead{(\degr)} & 
\colhead{(mas)} & 
\colhead{(\degr)} & 
\colhead{$\left(\frac{\mu {\rm Jy}}{{\rm bm}}\right)$} & 
\colhead{$\left(\frac{\mu {\rm Jy}}{{\rm bm}}\right)$} &
\colhead{0.5} & 
\colhead{0.6} & 
\colhead{0.7} & 
\colhead{0.8} & 
\colhead{0.9} & 
\colhead{1.0} \\
\colhead{(1)} & \colhead{(2)} & \colhead{(3)} & \colhead{(4)} & \colhead{(5)} & \colhead{(6)} & \colhead{(7)} & \colhead{(8)} & \colhead{(9)} & \colhead{(10)} & \colhead{(11)} & \colhead{(12)} & \colhead{(13)} & \colhead{(14)}
}
\startdata
SR 4      & D11  & \phn84 & \phn\phn4 & \phn84 & $+$105       & 127    & 35 & 118    & 127    & 138    & 150    & 178    & 230 \\
RU Lup    & D29  & 172    & \phn55    & 180    & $+$157       & \phn54 & 19 & \phn55 & \phn60 & \phn68 & \phn75 & \phn86 & 130 \\
Elias 20  & D25  & 120    & 280       & 178    & \phn$-$24    & \phn55 & 18 & \phn60 & \phn65 & \phn72 & \phn81 & \phn88 & 120 \\
Sz 129    & D41  & 192    & 242       & 226    & \phn\phn$+$1 & \phn44 & 16 & \phn44 & \phn49 & \phn56 & \phn63 & \phn72 & 160 \\
Sz 129    & D64  & 370    & 300       & 390    & \phn$-$53    & \phn51 & 15 & \phn53 & \phn56 & \phn60 & \phn68 & \phn82 & 150 \\
HD 143006 & D22  & 103    & 331       & 103    & \phn$-$73    & \phn70 & 25 & \phn71 & \phn79 & \phn88 & \phn97 & 111    & 220 \\
HD 143006 & D51  & 326    & \phn97    & 338    & $+$161       & \phn78 & 23 & \phn85 & \phn92 & \phn98 & 107    & 120    & 185 \\
GW Lup    & D74  & 446    & \phn\phn5 & 486    & $+$129       & \phn46 & 16 & \phn55 & \phn61 & \phn68 & \phn77 & \phn91 & 205 \\
Elias 24  & D57  & 341    & 137       & 394    & \phn\phn$-$2 & \phn55 & 13 & \phn62 & \phn68 & \phn75 & \phn82 & \phn91 & 175 \\
HD 163296 & D48  & 488    & 157       & 532    & \phn$+$57    & 118    & 36 & 124    & 135    & 147    & 162    & 188    & $>$250\phn\phd \\
HD 163296 & D86  & 591    & 226       & 861    & \phn\phn$-$2 & 104    & 31 & 112    & 124    & 133    & 143    & 156    & 235 \\
AS 209    & D61  & 428    & 347       & 520    & $-$173       & \phn52 & 16 & \phn56 & \phn60 & \phn68 & \phn77 & \phn86 & 175 \\
AS 209    & D97  & 689    & 330       & 814    & $-$159       & \phn65 & 15 & \phn70 & \phn77 & \phn83 & \phn89 & \phn97 & 155 \\
\enddata
\tablecomments{(1) Target name; (2) Gap designation from \citet{dsharp2}; (3) and (4) Residual peak location in sky-frame coordinates (projected radial distance from the disk center and position angle E of N); (5) and (6) Residual peak location in disk-frame polar coordinates (following the azimuth convention of \citealt{dsharp2}); (7) Peak residual brightness; (8) Standard deviation of pixel values in the gap; (9)--(14) Minimum mock CPD flux density that is recovered in 50, 60, 70, 80, 90, and 100\%\ (respectively) of the injection trials (based on spline interpolations of the profiles in Figure \ref{fig:recovs}).}
\end{deluxetable*}

As a guide to interpret these profiles, consider an over-simplified approximation where the non-CPD residuals are pure thermal noise (i.e., the {\tt frank} modeling is perfect, and there is no phase noise or imaging artifacts).  In that scenario, realizations of mock CPD injections result in point-like residuals with peak intensities consistent with random draws from a Gaussian distribution that has a mean equivalent to the injected flux density, $F_{\rm cpd}$, and a standard deviation that reflects the ``local" noise, RMS$_{\rm gap}$ (defined as the RMS of pixel values within the search annulus of the residual map).\footnote{A note on nomenclature and units: for the point-like residuals we assume here, the flux density (in $\mu$Jy) is equivalent to the peak intensity (in $\mu$Jy / bm) since the solid angle containing the signal is by default equivalent to the beam area.}  If a realization generates a mock CPD signal brighter than the {\it actual} peak residual in the search annulus, the recovery is successful.  Therefore, the 50\%\ recovery fraction occurs where the actual peak residual is $\approx F_{\rm cpd}$.  

Despite the assumptions being technically invalid, the above approximation is a reasonable reflection of the actual recovery outcomes.  Table \ref{table:resids} lists the locations and intensities for the actual peak residuals in both the sky-frame and disk-frame coordinate systems, along with the dispersions in the search annuli (RMS$_{\rm gap}$) and some CPD flux densities that correspond to representative recovery fractions (interpolated from the profiles in Figure \ref{fig:recovs}).  The measured RMS$_{\rm gap}$ values are generally $\sim$1.5--2$\times$ higher than found for a large, empty area (the RMS in Table \ref{table:data}), primarily due to lingering non-axisymmetric residuals from the disk modeling.  However, especially in the cases where there are few independent resolution elements sampling the search annulus (i.e., for narrow gaps and/or gaps located at small radii), the pixel distributions are not Gaussian (and strongly covariant) and therefore also tend to artificially inflate RMS$_{\rm gap}$. 

We estimated the ``false positive" probability of a recovery that is actually a thermal noise peak (i.e., drawn from a Gaussian with mean zero and dispersion RMS$_{\rm gap}$) and not the injected CPD signal as a function of $F_{\rm cpd}$.  There are two factors to consider: the probability of the peak falling within $\Delta R$ (i.e., the areal ratio of the recovery region and the gap search annulus), and the probability of the peak being brighter than $F_{\rm cpd}$.  The false positive recovery fraction can be approximated as the product of these factors, 
\begin{equation}
    \begin{aligned}
        &\approx \left[\frac{{\rm max}(\frac{\pi \langle \Delta R\rangle^2}{\Omega_{\rm bm}}, 1)}{4 \pi \sigma_{\rm gap} r_{\rm gap} \Omega_{\rm bm}^{-1}}\right] 
         \left[\frac{1}{2}\,{\rm erfc}\left(\frac{F_{\rm cpd}}{2 \, {\rm RMS}_{\rm gap}}\right)\right],
    \label{eq:false_positive}
    \end{aligned}
\end{equation}
where erfc is the complementary error function, $\langle \Delta R \rangle$ is the average distance threshold from Equation (\ref{eq:crit}) for each discrete flux density value, and $\Omega_{\rm bm}$ is the solid angle of the synthesized beam.\footnote{If each pixel were independent, we could write the first factor in Equation \ref{eq:false_positive} more simply as $\langle \Delta R \rangle^2 / 4 \sigma_{\rm gap} r_{\rm gap}$.  However, the version in the text reflects the spatial covariances in the residual images imparted by PSF convolution.  The differences between the two approximations are negligible for the cases presented here.}  This kind of false positive from random noise peaks is unimportant in our analysis, typically accounting for $\ll$\,1\%\ of the recoveries in the lowest flux density bins ($F_{\rm cpd} \lesssim {\rm RMS}_{\rm gap}$).  The only exception is for the narrow gap D11 in the SR 4 disk, where the false positive rate is consistent with the recovery rate ($\sim$few percent) at $F_{\rm cpd} \le 40$ $\mu$Jy.  

\begin{figure}[t!]
\includegraphics[width=\linewidth]{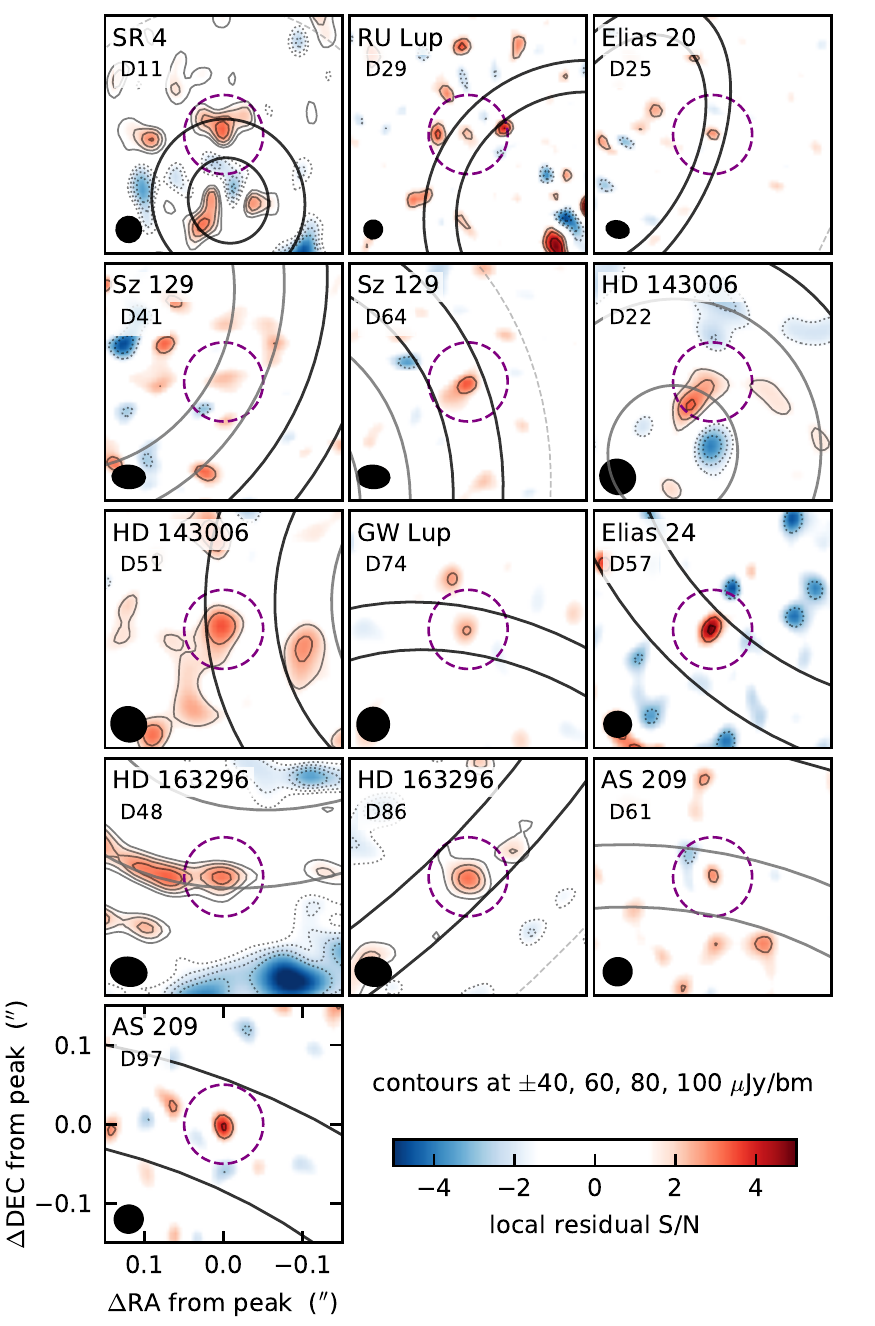}
\caption{Zoomed-in views of the purple dashed squares marked in the residual maps of Figure \ref{fig:resid_maps}, showing the regions immediately surrounding the actual residual peaks in each gap.  The annotations are as in Figure \ref{fig:resid_maps}; purple dashed circles mark the peaks.  The residual S/N scale is different here; the RMS$_{\rm gap}$ values in Table \ref{table:resids} are used to normalize the residual images, with a compressed S/N scale to emphasize the local background variations.  Contours are drawn at 40, 60, 80, and 100 $\mu$Jy bm$^{-1}$ (and their negative complements as dotted contours) to mark an absolute emission scale.     
\label{fig:peaks}
}
\end{figure}

The underlying causes of the {\it unsuccessful} recoveries in this adopted methodology are simply the actual residual peaks in the data.  These peaks are shown in the zoomed-in residual S/N maps of Figure \ref{fig:peaks}, and have their properties catalogued in Table \ref{table:resids}.  These peaks can be attributed to various origins, including noise, features associated with the circumstellar disk, or genuine CPDs.  As we noted in Section \ref{sec:disk_removal}, the gaps in the SR 4, HD 143006, and HD 163296 disks exhibit residual peaks that seem linked to the local non-axisymmetric behaviors of their circumstellar disks.  Despite these being perhaps the most compelling targets for planet-disk interactions, differentiating between these measured features and any point-like residuals that could be associated with CPDs will remain a challenge without deeper observations and improved disk emission models.  

The remaining cases suggest marginal residual peaks, often with a local ${\rm S/N}{\approx}3$ in their search annuli.  For the narrow gaps in the RU Lup, Elias 20, Sz 129, and GW Lup disks, peaks with such ``significance" are not necessarily expected (assuming the noise distribution is Gaussian) for search annuli with areas comparable to only $\sim$30--60 synthesized beam solid angles.  However, a close examination of the residuals in adjacent zones suggest that these are indeed consistent with local noise peaks.  The peaks in the Elias 24 D57 and AS 209 D97 gaps are slightly higher S/N cases that merit some follow-up.  We speculate that the peak in AS 209 D97 is probably associated with the very faint ring identified near the gap center \citep{dsharp8,dsharp2}.  We do not find any obvious residual signal near the location of the Elias 24 planet candidate identified by \citet{jorquera21}, nor in the vicinities of the CO kinematic perturbations in the HD 143006, HD 163296, GW Lup, or Sz 129 disks identified by \citet{pinte20}.     

These residual peaks are considered candidate CPDs with marginal confidence until they can be pursued with more sensitive measurements.  Since their peak intensities are similar to the $F_{\rm cpd}$ values that are recovered for 50\%\ of the injection--recovery experiments (Table \ref{table:resids}), we adopted the latter as a homogenized approximation for the ``upper limits" on CPD flux densities.

\section{Discussion} \label{sec:discussion}

\subsection{Limits on CPD Masses}

These derived upper limits on the mm continuum flux densities can be translated into analogous constraints on CPD (dust) masses.  We largely followed the approach of \citet{isella14,isella19}, although comparable results were found based on the methods of \citet{zhu18}.  The simplified model for the CPD emission is described in Appendix \ref{app:CPD_mass}.  This model permits an estimate of $M_{\rm cpd}$, defined as the mass corresponding to the flux density upper limit (the $F_{\rm cpd}$ with a 50\%\ recovery probability using the analysis described in Section \ref{sec:CPDs}; see column [9] in Table \ref{table:resids}).  We can alternatively think of $M_{\rm cpd}$ as the ``minimum detectable" mass, given the DSHARP data and our adopted assumptions and search methodology.  These mass limits were calculated as a function of the two key unknowns: the planet mass, $M_p$, and the CPD radius (in units of the Hill radius), $R_{\rm cpd}/R_{\rm H}$.

The various other parameters of the CPD model were either fixed (see Appendix \ref{app:CPD_mass}) or crudely explored around representative boundary values.  A more detailed discussion of the expected uncertainties associated with the `fixed' choices is available in Appendix \ref{app:CPD_mass}.  The CPD model densities are defined by the key input ($M_p$, $R_{\rm cpd}$) and output ($M_{\rm cpd}$) parameters.  The temperatures are based on approximations for irradiation heating by the planet and the host star (and local disk), as well as viscous heating from accretion.  The irradiation contributions assumed planetary evolution models and an approximation for the local disk temperatures \citep[e.g.,][]{chiang97}.  We considered two cases for the viscous heating.  In the first, accretion is not a significant contributor to the CPD thermal structure.  We assumed a constant accretion rate for the planet ``lifetime" ($t_p$), $\dot{M}_p \approx M_p / t_p$, with $t_p{=}1$ Myr (the mean age of the host stars; \citealt{dsharp1}).  \fix{This is roughly consistent with the $\dot{M}_p$ limits derived from H$\alpha$ non-detections in recent direct imaging CPD searches \citep[e.g.,][]{cugno19,hashimoto20,zurlo20}.}  In the second case, we scaled up that rate 100$\times$ to simulate an active accretion phase with a thermal contribution more comparable to (or larger than) irradiation.  

Moreover, we also explored two cases for the properties of the emitting dust grains in these model CPDs, as characterized by their absorption opacities ($\kappa_\nu$) and albedos ($\omega_\nu$).  In one, we assumed scattering is negligible ($\omega_\nu \approx 0$) and $\kappa_\nu = 2.4$ cm$^2$ g$^{-1}$ at the DSHARP data frequency (240 GHz).  This is more or less the standard assumption used so far in the literature.  The value for $\kappa_\nu$ is consistent with the classic \citet{beckwith90} opacity prescription.  In another case, we explored the effects of strong scattering ($\omega_\nu = 0.9$) for the same $\kappa_\nu$, more appropriate when most of the mass is concentrated in particles with sizes comparable to the observing wavelength \citep{dsharp5,zhu19}.

\begin{figure*}
\includegraphics[width=\linewidth]{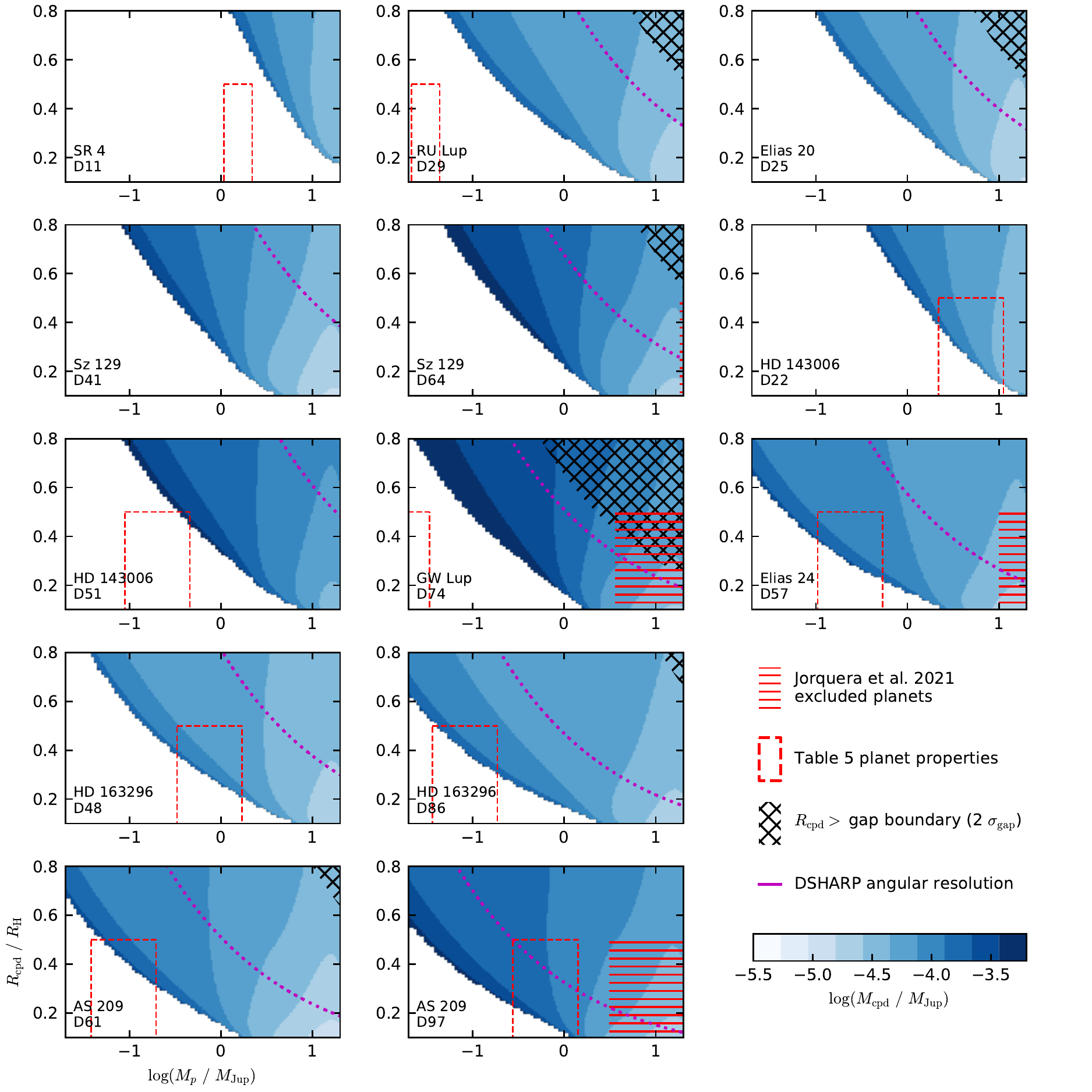}
\caption{The minimum detectable CPD dust masses as a function of the host planet mass and CPD radius (in Hill sphere radius units) for each target gap of interest, in this case presuming no scattering ($\omega_\nu \approx 0$) and modest accretion rates ($\dot{M}_p \approx M_p / {\rm 1 \, Myr}$).  The regions above the dotted magenta curves correspond to CPDs larger than the resolution element.  The black hatched regions correspond to CPDs larger than the gap size ($R_{\rm cpd} > 2 \sigma_{\rm gap}$).  The former technically violates our assumptions; the latter is unphysical.  For the empty white regions, no amount of mass is sufficient to produce the corresponding mm flux density, given our assumptions (i.e., the optical depths are saturated).  The red hatched regions correspond to the properties ruled out by the direct imaging searches by \citet{jorquera21}.  The red dashed rectangles mark the $M_p$ regions inferred with the simulations and methodology of \citet{dsharp7} (see Table \ref{table:planets}).  \change{{\it The full Figure Set is appended after the main text of the article.}}
\label{figset:Mcpd_lims}
}
\end{figure*}

The four components of Figure Set \ref{figset:Mcpd_lims} illustrate how the CPD mass upper limits for each gap vary as a function of $M_p$ and $R_{\rm cpd}/R_{\rm H}$, corresponding to each of the four distinct sets of assumptions outlined above.  Each panel also shows a magenta dotted curve that marks the (radial) resolution of the data (HWHM), and a black hatched region that denotes where the CPD size would be larger than twice the Gaussian standard deviation that describes the gap width ($2\sigma_{\rm gap}$; see Table \ref{table:gaps}).  Models that lie above the magenta curve violate our assumption of point-like CPD emission, and models within the black hatched region are ``unphysical" in the sense that the CPD should not extend beyond the gap boundaries.\footnote{As we noted in Section \ref{sec:modeling_results}, a Gaussian is not always the best representation of the gap emission distribution in the {\tt frank} models.  We hatched the region starting at twice the derived width (gap standard deviation) as a more conservative measure of the gap boundary to ensure that this ambiguity is not misleading.}  The derived $M_{\rm cpd}$ limits are in the range 0.001--0.2 M$_\oplus$ (roughly 10\%\ of a lunar mass to a Mars mass).  There is only a modest dependence on $R_{\rm cpd}/R_{\rm H}$ (see \citealt{isella19}), but we see clear decreases in the limits for larger $M_p$ and $\dot{M}_p$.  The latter behavior is caused by the associated heating: these CPD models are hotter, so less mass is required to produce a given flux.

Adjustments to the $M_{\rm cpd}$ limits for different dust albedos are more subtle (and still incomplete; e.g., scattering should also impact the CPD temperature structure, but this is not considered).  Generally, slightly lower $M_{\rm cpd}$ limits are found for higher $\omega_\nu$.  The cause of that behavior is complicated, but related to the variation of the dust optical depths.  In the optically thin limit, albedo has negligible effects.  And in the purely optically thick case, more scattering (higher $\omega_\nu$) means less emission \citep[e.g.,][]{zhu19}.  For the same physical parameters (temperatures and densities), a higher albedo means higher optical depths.  Therefore, in the general case that spans from high to low optical depths with increasing radius, the transition from thick to thin occurs at larger radii for higher albedo.  For our assumptions, this usually means higher flux densities for higher $\omega_\nu$, and therefore correspondingly lower $M_{\rm cpd}$ limits.

For low $M_p$ and $R_{\rm cpd}$ (the empty white regions in the lower left corners of these plots), the flux upper limits cannot be reproduced for {\it any} CPD mass with our adopted model assumptions.  These swaths are larger for models that make disks colder or scattering more prevalent.  In these regions of parameter-space, the measured flux upper limits are higher than for a scenario where the entire CPD is optically thick: the data are not sensitive enough to find CPDs with these assumed properties.  

To help contextualize these CPD mass limits, Figure Set \ref{figset:Mcpd_lims} highlights two sectors of parameter-space: the red hatched regions mark the $M_p$ ranges excluded (at 50\%\  probability) by the direct imaging measurements of \citet{jorquera21} (for the AS 209 D97, Elias 24 D57, and GW Lup D74, and Sz 129 D64 gaps), and the dashed red outlines correspond to the $M_p$ ranges predicted by the method of \citet{dsharp7}.  These latter sectors are based on comparisons of the relative gap widths ($\Delta$; see Eq.~21 in \citealt{dsharp7}) inferred from the {\tt frank} model brightness profiles (see Appendix \ref{app:SBprof}) and a suite of hydrodynamics simulations (both convolved with the same Gaussian kernel of width $0.06 \, r_{\rm gap}$), assuming a dust population with a maximum particle size of 1 mm (commensurate with the adopted $\kappa_\nu$) and a viscosity coefficient $\alpha{=}0.001$.  The corresponding $M_p$ (along with $\Delta$ and $a_p$) are compiled in Table \ref{table:planets}, with uncertainties estimated as described by \citet{dsharp7}.  For both the \citet{jorquera21} constraints and the \citet{dsharp7}-based masses, we highlighted the regions bound by $R_{\rm cpd}/R_{\rm H}{\approx}0.1$--0.5 as representative of theoretical expectations \citep[see][]{martin11}.  

\begin{deluxetable}{l l | c c | c}[t!]
\tablecaption{Estimated Planet Properties \label{table:planets}}
\tablehead{
\colhead{Disk} & 
\colhead{Gap} &
\colhead{$a_p$ (au)} & 
\colhead{$\log{(M_p / M_{\rm Jup})}$} & 
\colhead{$\Delta$} \\
\colhead{(1)} & \colhead{(2)} & \colhead{(3)} & \colhead{(4)} & \colhead{(5)}
}
\startdata
SR 4      & D11  & 10.7 & \phn\phd0.20 $^{+0.14}_{-0.17}$ & 0.44    \\ 
RU Lup    & D29  & 28.3 & $-$1.50 $^{+0.14}_{-0.17}$      & 0.14    \\
Elias 20  & D25  & 25.0 & $-$1.89 $^{+0.14}_{-0.17}$      & 0.12    \\
Sz 129    & D41  & 38.9 & $-$2.17 $^{+0.22}_{-0.16}$      & 0.16    \\
Sz 129    & D64  & 60.2 & \nodata$^{\dagger}$             & \nodata \\
HD 143006 & D22  & 23.4 & \phn\phd0.84 $^{+0.21}_{-0.50}$ & 0.70    \\
HD 143006 & D51  & 52.6 & $-$0.55 $^{+0.21}_{-0.50}$      & 0.22    \\
GW Lup    & D74  & 75.2 & $-$1.68 $^{+0.21}_{-0.50}$      & 0.14    \\
Elias 24  & D57  & 58.4 & $-$0.48 $^{+0.21}_{-0.50}$      & 0.30    \\
HD 163296 & D48  & 49.5 & \phn\phd0.02 $^{+0.21}_{-0.50}$ & 0.33    \\
HD 163296 & D86  & 85.9 & $-$0.94 $^{+0.21}_{-0.50}$      & 0.15    \\
AS 209    & D61  & 61.7 & $-$0.92 $^{+0.21}_{-0.50}$      & 0.21    \\
AS 209    & D97  & 96.8 & $-$0.06 $^{+0.21}_{-0.50}$      & 0.42    \\
\enddata
\tablecomments{(1) Target name; (2) Gap designation from \citet{dsharp2}; (3) estimated orbital semimajor axis, based on the derived distance (see Table \ref{table:data}) and $r_{\rm gap}$ (see Table \ref{table:gaps}; note, these are not perfect matches for the designations in column [2] due to both the distance adjustments based on {\it Gaia} EDR3 and the modified approach for measuring gap centers); (4) estimated masses and uncertainties, based on the \citet{dsharp7} approach and assuming a viscosity coefficient $\alpha = 0.001$ and a maximum particle size of 1 mm (see text); and (5) relative gap width, as defined by \citet{dsharp7} (their Equation 21). \\
$^{\dagger}$ The Sz 129 D64 gap is too narrow to reliably measure a planet mass using the \citet{dsharp7} method.
}
\end{deluxetable}

Generally, the mass limits derived here are similar to theoretical expectations for the CPDs predicted in simulations of gravitationally unstable disks \citep{stamatellos15} or the formation of the Galilean satellites \citep{canup02}.  For the SR 4, HD 143006, Elias 24, HD 163296, and AS 209 (D97) disk gaps, the \citet{dsharp7}-based predictions imply $M_{\rm cpd} / M_p \lesssim 1$--10\%\ (in some cases $\lesssim$~0.01\%\ or up to $\sim$60\%), presuming a global gas-to-dust ratio of 100.  In the other cases, the flux limits are above the expectations for fully optically thick CPDs; limits on $M_{\rm cpd}$ are unavailable if the planet masses in Table \ref{table:planets} are appropriate.

We should make an explicit reminder that these data and our analysis are only sensitive to CPD dust masses, and our opacity assumptions are specifically relevant for particles with $\sim$mm sizes.  However, smooth CPD density (or, rather, pressure) structures like those assumed in our simple model are expected to facilitate the rapid inward radial drift of such particles \citep[e.g.,][]{shibaike17}.  If there is no pressure modulation to ``trap" those particles \citep{drazkowska18,batygin20,szulagyi21}, a CPD could have a considerable total mass (in gas and small particles) but produce very little mm continuum emission.  In that sense, ALMA observations might only be sensitive to the small grains still dynamically coupled to the gas, or to CPDs that have their own substructures.

\subsection{Empirical Context for CPD Limits}

\begin{figure}[t!]
\includegraphics[width=\linewidth]{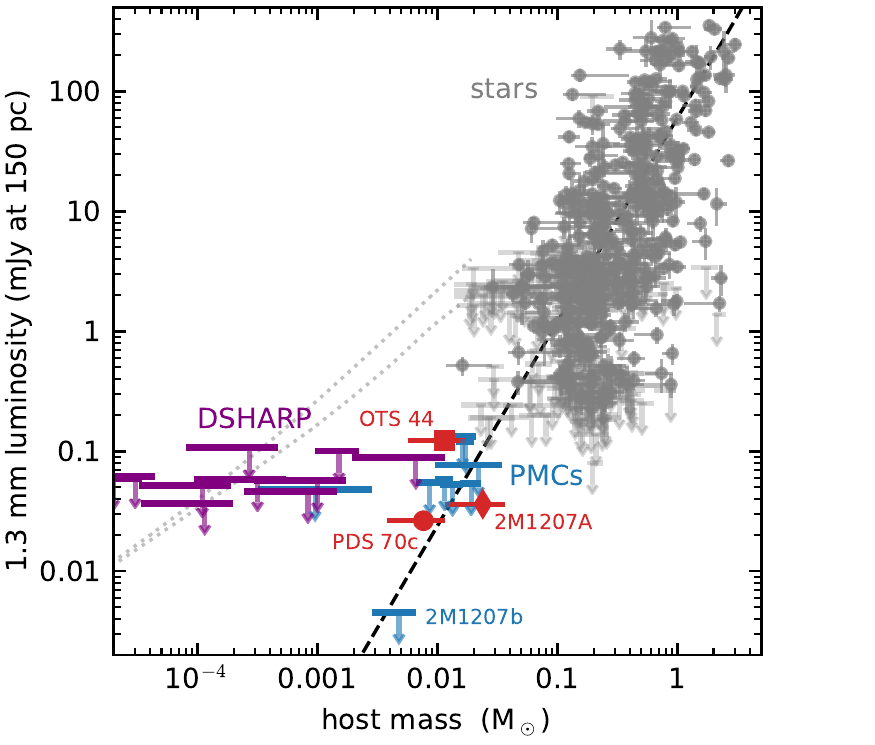}
\caption{The 1.3 mm continuum luminosity (flux normalized to a common distance of 150 pc) for disk material as a function of the stellar or planetary host mass; adapted from \citet{andrews20}.  Upper limits are marked with downward-directed arrows.  When 1.3 mm data were unavailable, the 0.8--0.9 mm measurements were scaled by the mean continuum spectrum $\nu^{2.2}$ \citep{andrews20,tazzari21}.  The black dashed line is the mean correlation derived from nearby circumstellar disks (in gray; \citealt{andrews13,ansdell16,pascucci16,barenfeld17,cieza19,williams19,akeson19}).  The datapoints for planetary mass companions (PMCs) are taken from \citet{macgregor17}, \citet{pineda19}, and \citet{wu20}, the 2M1207 data from \citet{ricci17}, the OTS 44 data from \citet{bayo17}, and the PDS 70c data from \citet{isella19}.  As a point of reference, the dotted gray curves show fully optically thick CPD models for the median properties of the DSHARP targets, $R_{\rm cpd}/R_{\rm H} = 0.3$, and the default (lower curve) and high (upper curve) accretion rate scenarios described above (assuming no scattering).  
\label{fig:lmm_mstar}
}
\end{figure}

Inferences of CPD mass constraints like those above should be considered with due caution: they still require many assumptions about physical properties where our knowledge is very limited.  As an alternative, we can examine a more empirical context for these CPD constraints by comparing them with other samples in the mm continuum luminosity--``host" mass domain, as illustrated in Figure \ref{fig:lmm_mstar}.  The upper limits derived in this article are comparable to those for planetary-mass companions (PMCs; \citealt{bowler15,macgregor17,ricci17,pineda19,wu20}), the isolated planet OTS 44 \citep{bayo17}, and the planet PDS 70c \citep{isella19}, although often for much lower ``host" masses \citep{dsharp7}.  While these limits rule out optically thick CPDs around the more massive putative planets (e.g., the gray dotted curves in Figure \ref{fig:lmm_mstar}), they would lie well above ($\gtrsim 10$-100$\times$) the luminosities expected if the correlation identified for circumstellar disks \citep[see the discussion by][]{andrews20} is extrapolated to lower host masses (black dashed line in Figure \ref{fig:lmm_mstar}).  Of course, since we do not yet understand the physical origin of that correlation, there is not necessarily a good reason to expect that it would extend indefinitely into the planetary mass regime. 

Perhaps a more appropriate comparison to the PMC disk searches would be an exploration of putative CPDs in the DSHARP sample located at larger distances than were considered here, outside the continuum emission boundary ($a_p \gtrsim R_{\rm out}$).  In one sense, searches in these regions -- where there are no non-axisymmetric residuals from the host disk present -- are much simpler.  Using the formalism described in Section \ref{sec:CPDs}, we have confirmed that upper limits (50\%\ recovery fractions) for point-like features are what we would naively expect, $\sim$3$\times$ the map RMS noise levels (listed in Table \ref{table:data}).  Those limits would be appropriate in the limiting case where efficient radial drift makes the distribution of emitting solids especially compact.  But in the more general case, the real challenge is that theoretical models \citep[e.g.,][]{martin11} suggest that CPDs at these distances will be spatially resolved with the DSHARP data: we would need to substantially modify the adopted search methodology to properly quantify flux limits for such CPDs.

One general assumption that we make is that CPDs will exhibit brighter mm emission if they have more massive planetary hosts, because there is more irradiation (and viscous) heating and (presumably) more mass.  The dynamical perturbations such massive planets induce on the structures of the circumstellar disks in which they are embedded are expected to be more pronounced than the narrow gaps observed in the DSHARP sample (i.e., see the $M_p$ values in Table \ref{table:planets}).  The large, cleared cavities of `transition' disks are nominally better search targets for these brighter CPDs around more massive planets \citep[e.g.,][]{salyk09,zhu11}.  Clearly, the PDS 70 system offers the most compelling support for such a strategy \citep{isella19,benisty21}.  However, it might also be the case that CPDs in transition disk cavities are relatively starved of the $\sim$mm-sized particles that emit most efficiently in the ALMA bands because the supply flow of those particles is throttled at the high-amplitude pressure maxima their host planets induce at the cavity edge \citep[e.g.,][]{pinilla12b,pinilla15b}.  Contrary to our simple expectations, this might imply that CPDs in transition disk cavities will only emit weak mm continuum radiation despite their higher temperatures and (potentially) higher gas masses.

\subsection{Future Directions}

In any case, clear and meaningful constraints on the mm emission from CPDs for most of the DSHARP sample will require more sensitive measurements.  This will likely mean a shift in strategy to target ALMA observations at higher frequencies, where the CPD is brighter (e.g., as advocated by \citealt{szulagyi18}).  That comes with practical obstacles (more noise and less time available in favorable observing conditions), but also physical challenges.  The local circumstellar disk emission will also be brighter at these frequencies, and could be considerably more complex.  At higher frequencies, the continuum emission traces smaller particles that are better coupled to the gas, which often results in emission gaps that are narrower and shallower \citep[e.g.,][]{tsukagoshi16,carrasco-gonzalez19,huang18,huang20,long20,macias21}.  Emission from such particles in the gaps would have higher optical depths, and could contribute to circumstellar extinction of CPD signals.  The search methodology advocated here would be beneficial with such data, although development of a more sophisticated recovery algorithm (rather than the simplistic peak/outlier identification that we have adopted) would likely be necessary.

It is especially striking that the disks which are predicted to host the most massive planets exhibit the most prominent non-axisymmetric residuals (HD 143006, SR 4, and HD 163296, according to Table \ref{table:planets}; also Elias 24 at lower levels, which is made more intriguing by the tentative companion found by \citealt{jorquera21}).  There is compelling indirect evidence for a perturber in the HD 143006 D22 gap \citep[e.g.,][]{ballabio21}, including a warp identified through extreme scattered light shadows \citep{benisty18}, the CO gas kinematics and continuum morphology at smaller radii \citep{dsharp10}, as well as the faint $m{=}1$ spiral identified here.  As we noted above, these more massive companions nominally exhibit brighter mm continuum emission from their (warmer) CPDs, but their robust detection is unfortunately much more difficult due to these asymmetries.  More than raw sensitivity, this sort of non-axisymmetric confusion limit could end up being the key obstacle in ``blind" searches for CPDs like the effort presented here.  It is possible that detectable CPD emission is present in these systems with the data used here, but we are incapable of differentiating it from asymmetric residuals that are not captured in the modeling.  

In future work that pushes to improved sensitivity, it will be important to develop more flexible models of the circumstellar disk emission to help track down these faint CPDs.  In some scenarios, particularly when the search region is confined by prior information, direct visibility modeling will outperform other approaches.  Perhaps the biggest advance in hunting for mm continuum CPD emission will be forthcoming direct imaging planet detections that enable such {\it targeted} searches and more cohesive explorations of the connections between planets, CPDs, and the disk substructures they create.

\section{Summary} \label{sec:summary}

We used the high resolution ALMA 1.25 mm continuum observations from the DSHARP survey to search for faint emission from circumplanetary material in the narrow gaps of circumstellar disks.  Our key findings from this effort are summarized as follows.  
\begin{itemize}
    \item We developed a prescription to mitigate contamination from the local circumstellar disk material, using the {\tt frank} software package \citep{jennings20}, and a methodology for statistically quantifying the sensitivity to point-like CPD emission using injection--recovery experiments.
    \item We found a few examples of pronounced asymmetric residuals in the target disks.  The most interesting case is a faint, one-armed spiral that traverses across all of the axisymmetric substructures in the HD 143006 disk, possibly driven by a companion in the innermost (D22) gap.
    \item There are a few peak residuals in these gaps that are  marginal CPD candidates; deeper observations (preferably at better resolution) would be required to establish confidence that they are not merely local noise peaks.  Upper limits on any CPD flux densities are 50--70 $\mu$Jy in most cases, rising to 110 $\mu$Jy in the few targets with lingering non-axisymmetric features within their gaps.
    \item If the gaps in these DSHARP disks are opened by giant planets with masses comparable to Jupiter, these constraints correspond to CPD (dust) mass upper limits of $\sim$0.001--0.2 M$_\oplus$.  Alternatively, if the planet masses are much lower (as is the prediction for some targets based on the hydrodynamics simulations of \citealt{dsharp7}), then considerably deeper observations may be required in future (sub-)mm continuum CPD searches.  Hopefully, those will be guided by direct imaging detections of the young planet hosts.
\end{itemize}

\acknowledgments 
We are especially grateful to Kees Dullemond and Xue-Ning Bai for their helpful comments on the draft manuscript, Ryan Loomis for his technical advice and kindly providing access to {\tt CASA} scripts, and Holly Thomas for her assistance facilitating computational efforts with the resources of the Smithsonian Radio Telescope Data Center.  S.A. and W.E. acknowledge support from the National Aeronautics and Space Administration under grant No.~17-XRP17$\_$2-0012 issued through the Exoplanets Research Program.  The analysis in this article uses the following ALMA datasets: 
\begin{itemize}
    \item ADS/JAO.ALMA \#2016.1.00484.L
    \item ADS/JAO.ALMA \#2013.1.00226.S
    \item ADS/JAO.ALMA \#2013.1.00366.S
    \item ADS/JAO.ALMA \#2013.1.00498.S
    \item ADS/JAO.ALMA \#2013.1.00601.S
    \item ADS/JAO.ALMA \#2015.1.00486.S
    \item ADS/JAO.ALMA \#2015.1.00964.S. 
\end{itemize}
ALMA is a partnership of ESO (representing its member states), NSF (USA), and NINS (Japan), together with NRC (Canada), MOST and ASIAA (Taiwan), and KASI (Republic of Korea), in cooperation with the Republic of Chile. The Joint ALMA Observatory (JAO) is operated by ESO, AUI/NRAO, and NAOJ.  This work used data from the European Space Agency (ESA) mission {\it Gaia} (\url{https://www.cosmos.esa.int/gaia}), processed by the {\it Gaia} Data Processing and Analysis Consortium (DPAC, \url{https://www.cosmos.esa.int/web/gaia/dpac}). Funding for the DPAC has been provided by national institutions, in particular the institutions participating in the {\it Gaia} Multilateral Agreement.

\facilities{ALMA}

\software{
{\tt CASA} \citep{mcmullin07}, 
{\tt numpy} \citep{numpy}, 
{\tt scipy} \citep{scipy}, 
{\tt matplotlib} \citep{matplotlib}, 
{\tt astropy} \citep{astropy},
{\tt frank} \citep{jennings20},
{\tt vis\char`_sample} (\url{https://github.com/AstroChem/vis\_sample}).
}

\begin{figure*}
\figurenum{8.1}
\includegraphics[width=\linewidth]{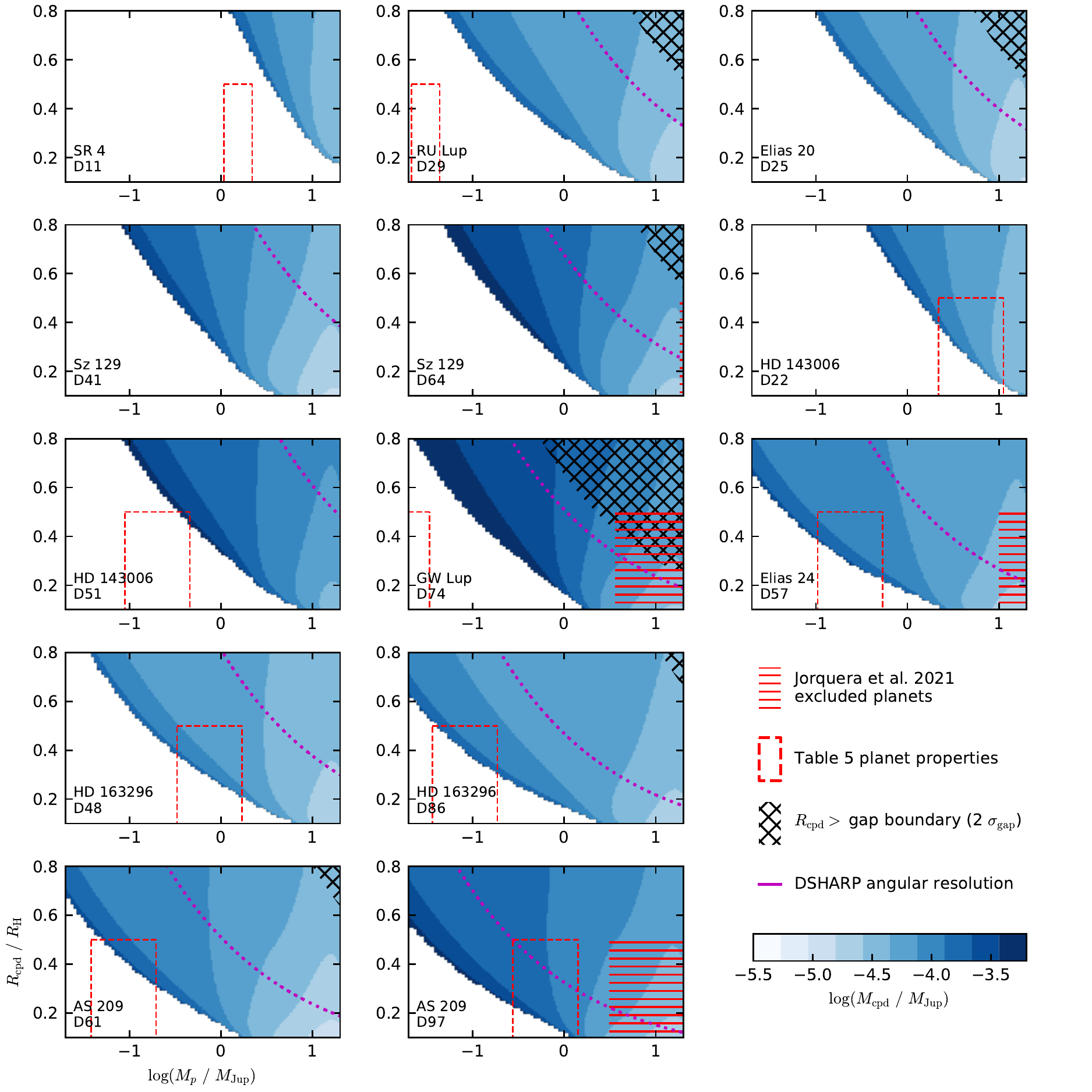}
\caption{The minimum detectable CPD dust masses as a function of the host planet mass and CPD radius (in Hill sphere radius units) for each target gap of interest, in this case presuming no scattering ($\omega_\nu \approx 0$) and modest accretion rates ($\dot{M}_p \approx M_p / {\rm 1 \, Myr}$).  The regions above the dotted magenta curves correspond to CPDs larger than the resolution element.  The black hatched regions correspond to CPDs larger than the gap size ($R_{\rm cpd} > 2 \sigma_{\rm gap}$).  The former technically violates our assumptions; the latter is unphysical.  For the empty white regions, no amount of mass is sufficient to produce the corresponding mm flux density, given our assumptions (i.e., the optical depths are saturated).  The red hatched regions correspond to the properties ruled out by the direct imaging searches by \citet{jorquera21}.  The red dashed rectangles mark the $M_p$ regions inferred with the simulations and methodology of \citet{dsharp7} (see Table \ref{table:planets}).
}
\end{figure*}

\begin{figure*}
\figurenum{8.2}
\includegraphics[width=\linewidth]{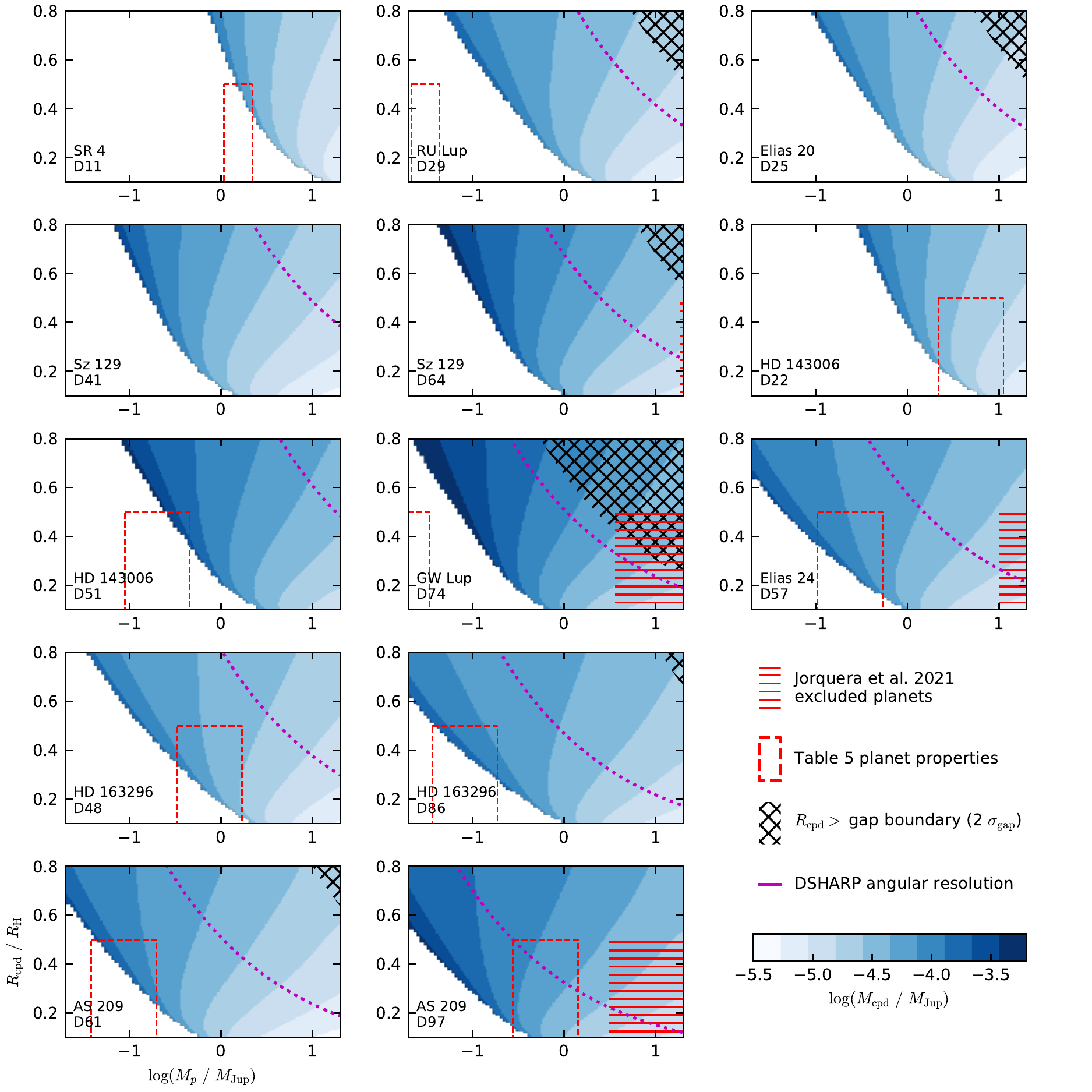}
\caption{As in Figure 8.1, but in the case of more vigorous accretion rates ($\dot{M}_p \approx 100 M_p / {\rm 1 \, Myr}$).
}
\end{figure*}

\begin{figure*}
\figurenum{8.3}
\includegraphics[width=\linewidth]{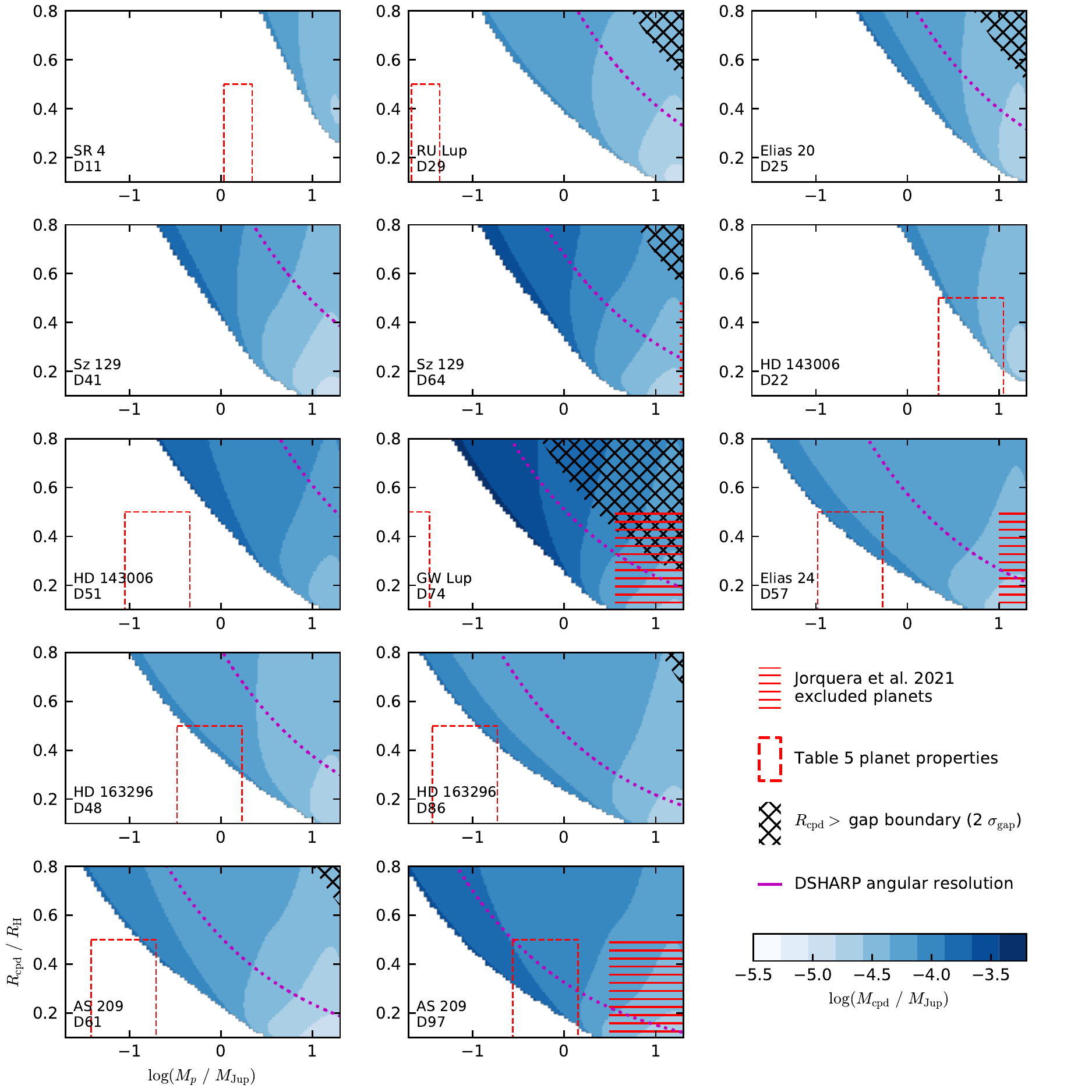}
\caption{As in Figure 8.1, but in the case where the emitting particles have high albedos ($\omega_\nu = 0.9$).  }
\end{figure*}

\begin{figure*}
\figurenum{8.4}
\includegraphics[width=\linewidth]{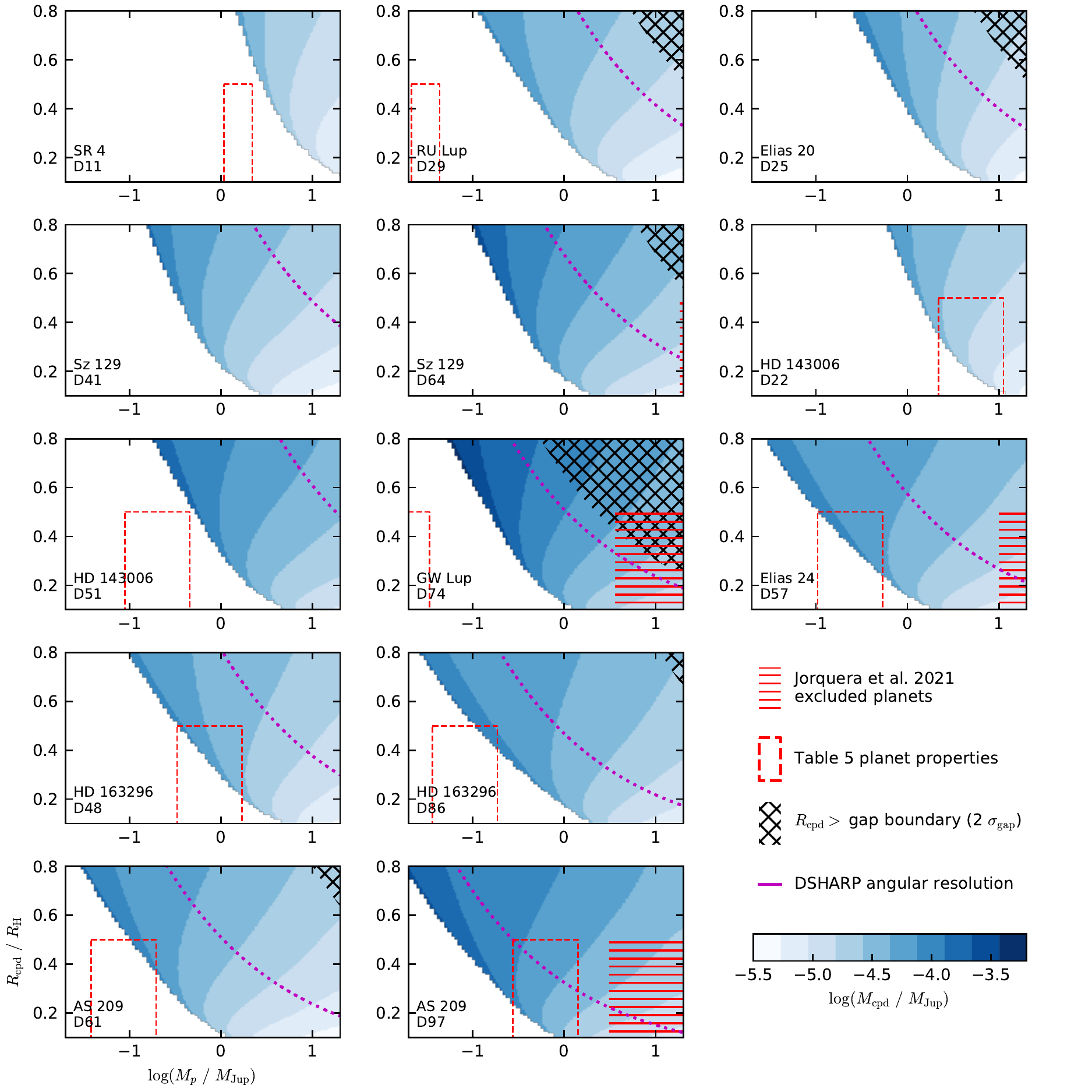}
\caption{As in Figure 8.1, but in the case where the emitting particles have high albedos ($\omega_\nu = 0.9$) and the disks have more vigorous accretion rates ($\dot{M}_p \approx 100 M_p / {\rm 1 \, Myr}$).
}
\end{figure*}

\clearpage

\appendix

\section{Notes on Asymmetric Residuals} \label{app:resid}

The modeling procedure outlined in Section \ref{sec:disk_removal} generally performs well in removing axisymmetric circumstellar disk emission that might contaminate the search for CPD emission.  Understanding the patterned morphologies of lingering residuals can be useful for revising the (fixed) geometric parameters of the disk and interpreting other potential sub-optimal assumptions made in the modeling.  Using simple empirical models, we offer some brief guidance here to help identify the origins of such ``artificial" residual features in future efforts (see also a similar discussion by \citealt{jennings21}).

We generated a series of synthetic datasets for this task.  In each case, we assumed a fixed radial brightness temperature profile (in the Rayleigh-Jeans limit)
\begin{equation}
    T_b(r) = 15 \, \left(\frac{r}{70 \, {\rm mas}}\right)^{-0.5} {\rm K}
\end{equation}
out to a radius of 0\farcs65, beyond which it decreases like $r^{-5}$.  Two narrow gaps were imposed on this profile, from 100--120 mas and 470--540 mas, where the base profile above was multiplied by 0.01 and 0.05, respectively.  This emission distribution in disk-frame polar coordinates ($r$, $\varphi$) was used to make a synthetic image on a sky-frame coordinate system,
\begin{equation}
    \begin{aligned}
        x_s &{=}r \sin{\varphi} \sin{{\rm PA}}{+}(r \cos{\varphi} \cos{i}{-}z_{\rm s} \sin{i}) \cos{{\rm PA}} \\
        y_s &{=}r \sin{\varphi} \cos{{\rm PA}}{-}(r \cos{\varphi} \cos{i}{-}z_{\rm s} \sin{i}) \sin{{\rm PA}}
    \end{aligned}
    \label{eq:proj_geom}
\end{equation}
where $z_{\rm s}$ is the vertical height of the emission surface.  In all of the analysis in the main text, and unless otherwise specified here, we have assumed $z_{\rm s} = 0$.  The center of the geometry specified in Equation (A2) could be offset from the image center by ($\Delta x$, $\Delta y$), with positive values corresponding to shifts to the E and N, respectively.  A synthetic visibility dataset was generated from the  Fourier transform of each image, sampled at the same spatial frequencies as the observations for the GW Lup disk, using the {\tt vis\char`_sample} software package.\footnote{\url{https://github.com/AstroChem/vis\_sample}}  

\begin{figure}[t!]
\includegraphics[width=\linewidth]{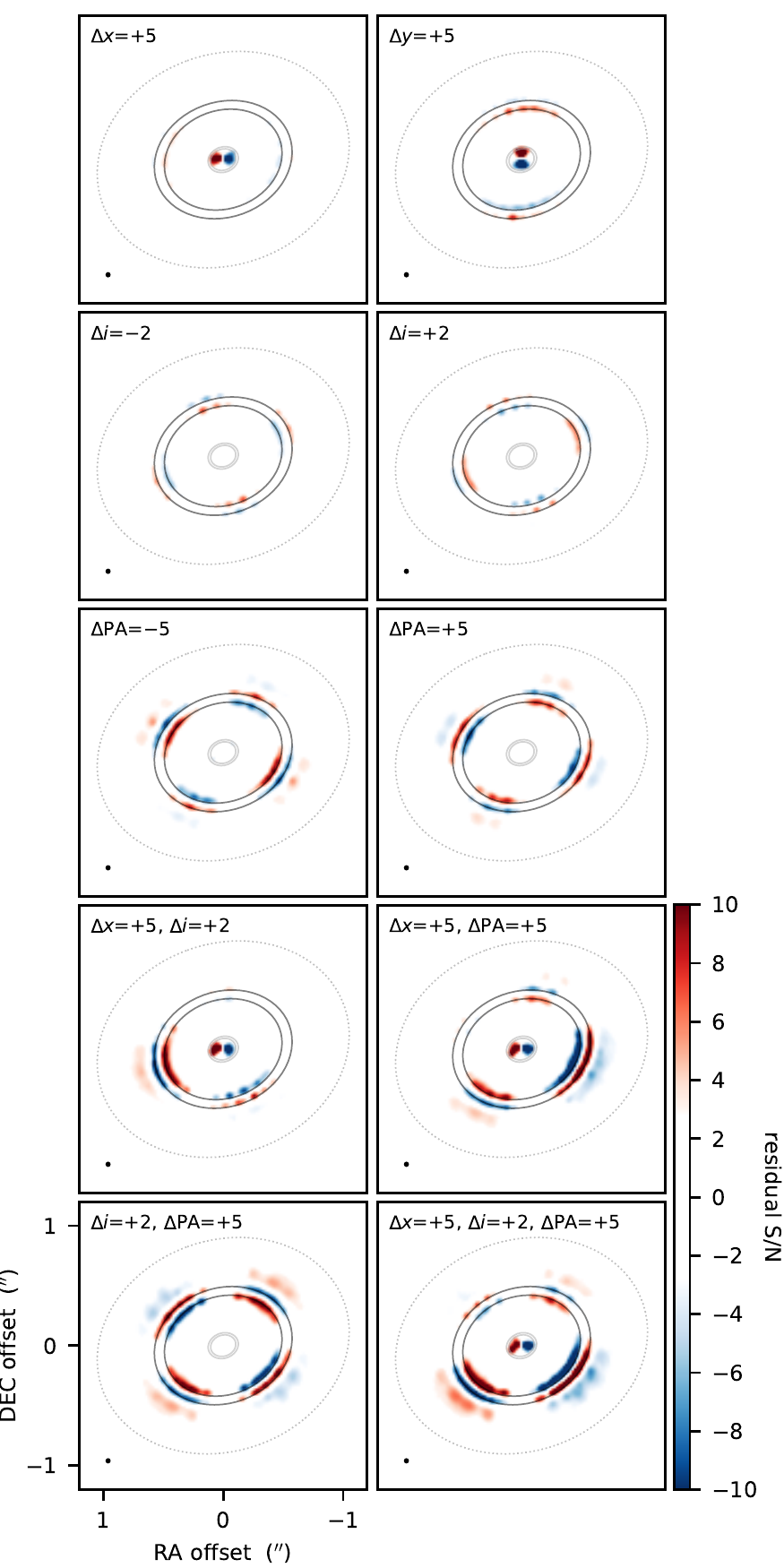}
\caption{Residual S/N images constructed from {\tt frank} modeling with the default geometric parameters in the text, where the data were generated with the deviations from the default parameters noted at the top left of each panel.  All annotations are the same as in Figure \ref{fig:resid_maps}.  The ``noise" in this case is set to the RMS value for the GW Lup disk (Table \ref{table:data}).  
\label{fig:geom1}
}
\end{figure}

\begin{figure}[t!]
\includegraphics[width=\linewidth]{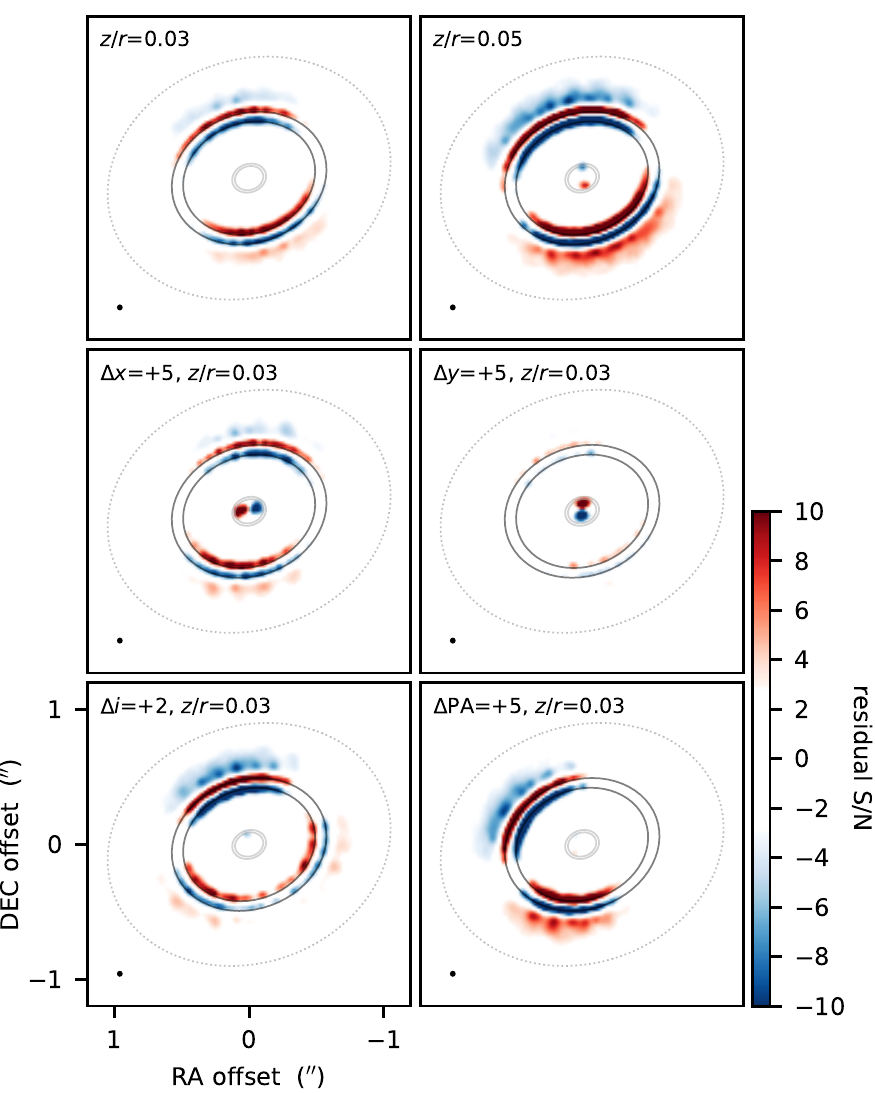}
\caption{Residual S/N images as in Figure \ref{fig:geom1}, but in this case associated with elevated emission surfaces.
\label{fig:geom2}
}
\end{figure}

These synthetic visibilities were then modeled with {\tt frank} following the same procedure outlined in Section \ref{sec:disk_removal}.  We assumed in this modeling the default geometric parameters ($\Delta x$, $\Delta y$) = (0 mas, 0 mas), $i = 35$\degr, PA = 110\degr, and $z_{\rm s} = 0$ mas (the latter a requirement for {\tt frank}).  However, the synthetic visibilities were generated for a sequence of deviations from these default values.  Figure \ref{fig:geom1} shows the imaged residual visibilities (as in Figure \ref{fig:resid_maps}) to illustrate the cases where the offsets, inclination, and/or PA have been slightly mis-assigned in the modeling with respect to the data.  These residual images demonstrate that even small offsets produce strong +/- residuals around the disk center, inclination mismatches are seen most prominently near ``edges" (of the outer gap in this case) as quadrupolar +/- residuals along the major and minor axes, and an incorrect PA assignment shows a similar behavior but rotated off-axes.  Combinations of these features are somewhat more difficult to disentangle, although the asymmetric residual morphologies generated when an offset is mis-specified are clear.  When optimizing the geometry, a decision on the disk center can usually be made first and fixed before the projection angles are explored in more detail.

Figure \ref{fig:geom2} shows the analogous residual behavior when considering elevated emission surfaces, characterized with constant aspect ratio $z_{\rm s}/r$, as might be expected in cases where continuum optical depths are high.  Indeed, it was behavior like the top panels in some initial modeling exploration that led us to exclude some of the more highly inclined disk targets in the DSHARP sample -- where such effects are most prevalent -- from the analysis presented here.  Modeling an elevated emission surface with a model that presumes an intrinsically flat morphology results in a pronounced, symmetric residual pattern along the minor axis.  This behavior can be misinterpreted with other methodologies as a spatial offset in the minor axis direction, but a careful examination of the residuals near the disk center can help distinguish the difference (see the middle right panel).  Mixing surface and other geometric effects can create complicated residual patterns (e.g., see the HD 163296 residuals).

\section{Model Treatment of Confined Azimuthal Asymmetries} \label{app:asym}

\begin{figure*}[ht!]
\includegraphics[width=\linewidth]{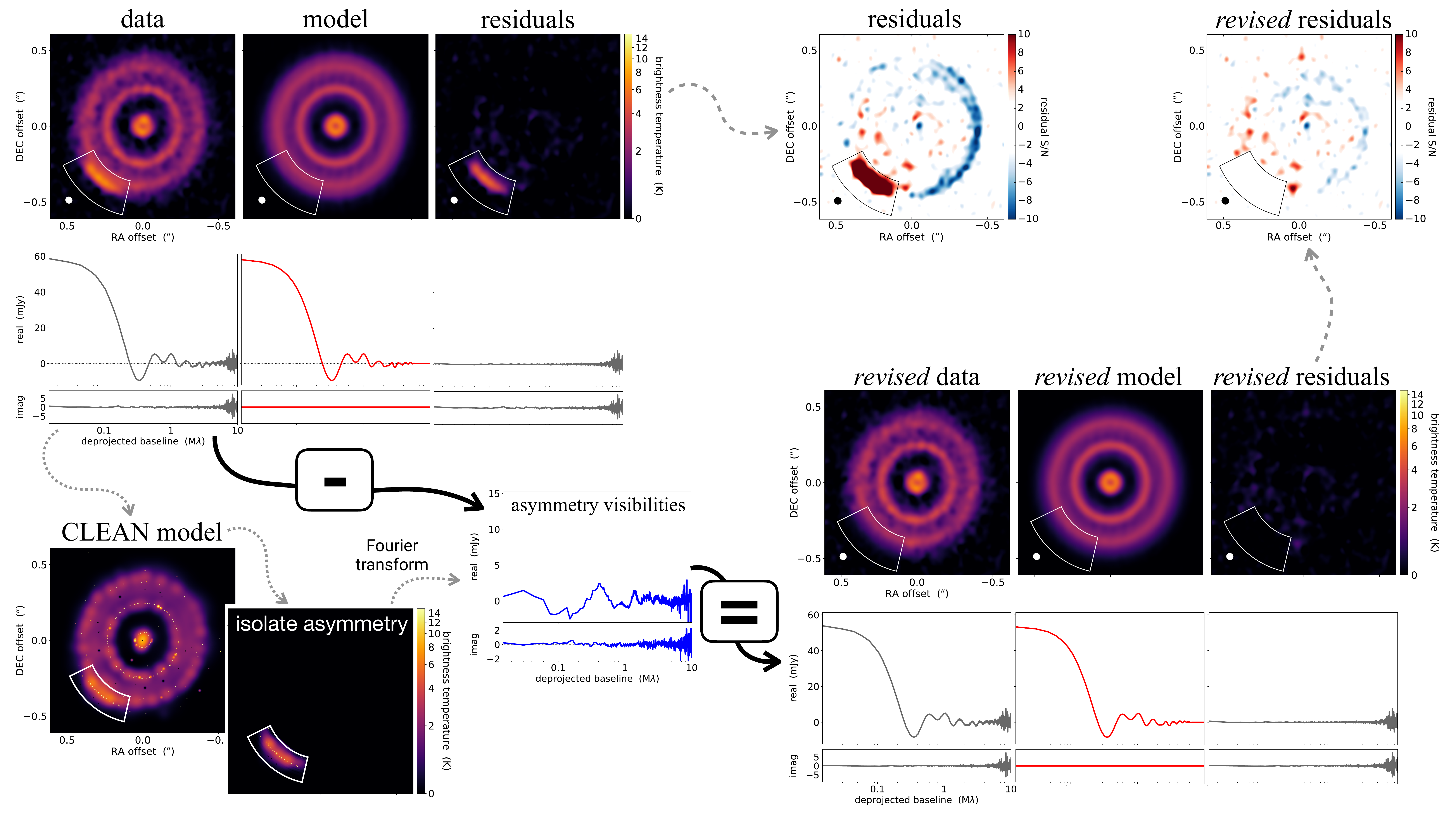}
\noindent\makebox[\linewidth]{\rule{\linewidth}{1pt}} \\ \\
\includegraphics[width=\linewidth]{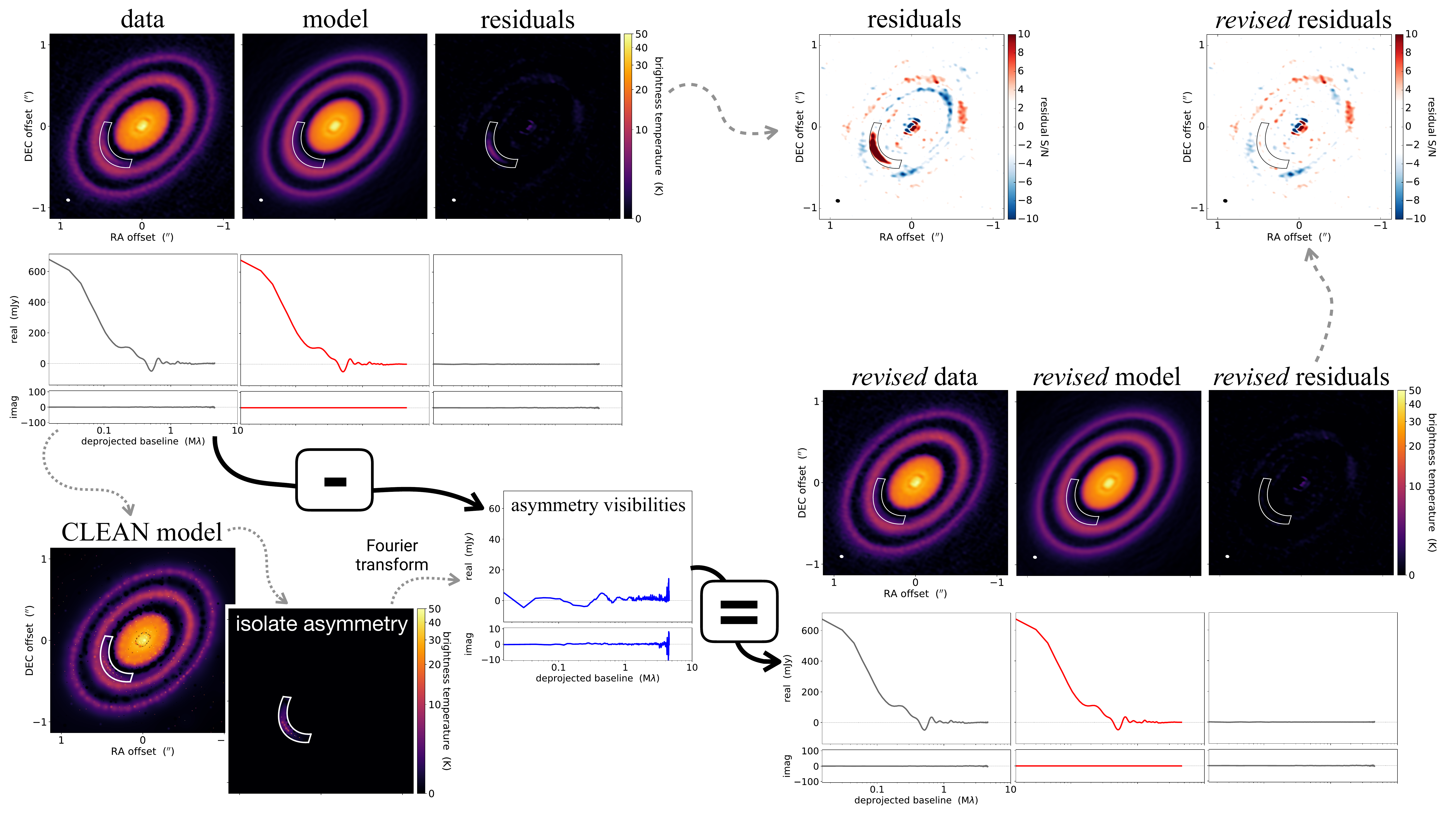}
\caption{Graphical overviews of the workflow for removing the confined azimuthal asymmetries (the ``arc" features confined to the regions outlined in each image) for the HD 143006 (top) and HD 163296 (bottom) disks.  The upper left panels show the effects of modeling the emission without treating the asymmetries, in both the images (as in Figures \ref{fig:dmrs1} and \ref{fig:dmrs2}) and visibilities (as in Figure \ref{fig:visprofs}).  The bottom left panels illustrate how the asymmetry models are constructed.  And the bottom right panels demonstrate the results of modeling the revised data, where the asymmetries have been removed.  The upper right panels directly compare the residual S/N maps (as in Figure \ref{fig:resid_maps}) for the uncorrected (with asymmetry) and revised (asymmetry removed) cases.   
\label{fig:asymm_workflow}
}
\end{figure*}

The HD 143006 and HD 163296 disks have pronounced, but spatially confined, `arc'-like azimuthal features in their emission distributions that present a challenge for the standard axisymmetric modeling methodology outlined in Section \ref{sec:disk_removal} \citep{dsharp9,dsharp10}.  Specifically, these asymmetries are sufficiently bright that ignoring them leads {\tt frank} to derive axisymmetric models that over-predict the emission at comparable radii, resulting in a pronounced negative residual at azimuths that lie well away from the asymmetry.  A demonstration of that effect is clear in the residual S/N images shown in the top right panels of Figure \ref{fig:asymm_workflow}.    

We designed a workaround that first revises the data visibilities by removing a simple model for the asymmetry before performing the {\tt frank} modeling described in Section \ref{sec:disk_removal}.  This asymmetry model was constructed first by spatially isolating the feature in the {\sc clean} model image, setting the {\sc clean} components outside a specific area associated with the asymmetry to zero.  The area of interest was selected manually by specifying radial and azimuthal boundaries in the disk plane.  For HD 143006, this corresponded to the annular arc spanning radii from 0.37 to 0\farcs60 and azimuths from 90 to 142\degr, following the \citet{dsharp2} azimuth convention (where 90\degr\ coincides with the major axis and azimuths increase clockwise on the sky).  For HD 163296, the radial and azimuthal boundaries were 0.48 to 0\farcs60 and 50 to 150\degr, respectively.  Next, the mean radial profile in the {\sc clean} model constructed from outside that region was subtracted from the asymmetry model image, leaving only the asymmetric contribution.  These steps are illustrated in the bottom left parts of Figure \ref{fig:asymm_workflow}.  

Next, the Fourier transform of the asymmetry model image was sampled at the observed spatial frequencies and then subtracted from the original data visibilities.  Finally, the resulting revised data visibilities were then modeled as described in Section \ref{sec:disk_removal}.  The entire process was iterated to settle on appropriate geometric parameters before finalizing the modeling outcomes.  

The full workflows for treating these special cases are illustrated together in Figure \ref{fig:asymm_workflow}.  A direct comparison of the residual S/N images in the upper right sections of these composite figures demonstrate a notable improvement in the fit quality achieved by first removing the confined azimuthal asymmetries.  However, it is interesting to see that these two cases with the most pronounced non-axisymmetric features still end up exhibiting considerable low-level asymmetric structure that is not captured by the modeling (e.g., see especially Figure \ref{fig:HD143_spiral}).  Presumably that is a distinctive, albeit subtle, clue to the origins of the substructures in these cases.  

\begin{figure*}[t!]
\includegraphics[width=\linewidth]{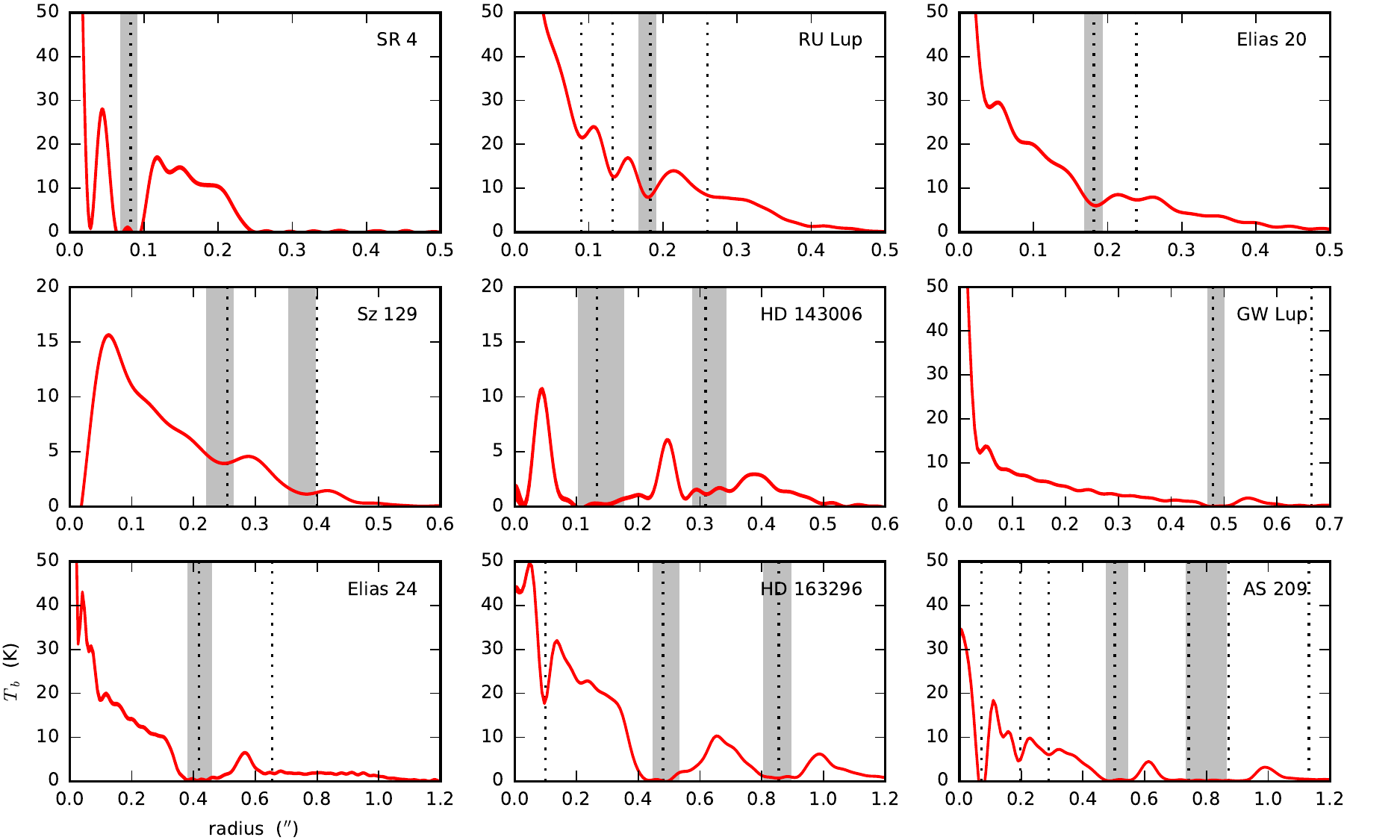}
\caption{The brightness temperature radial profiles for the circumstellar disk emission derived in Section \ref{sec:disk_removal} using the {\tt frank} package (assuming the Rayleigh-Jeans approximation).  Gray bands mark the gap annuli ($r_{\rm gap} \pm \sigma_{\rm gap}$); dotted lines denote the gap centers measured in the image plane by \citet{dsharp2}.  As a reminder, the gap feature at $\sim$0\farcs8 in the AS 209 disk (D97) was treated as two gaps separated by a faint ring by \citet{dsharp2}; see the note in Table \ref{table:gaps}.  
\label{fig:Tbprofs}
}
\end{figure*}

\begin{figure*}[t!]
\includegraphics[width=\linewidth]{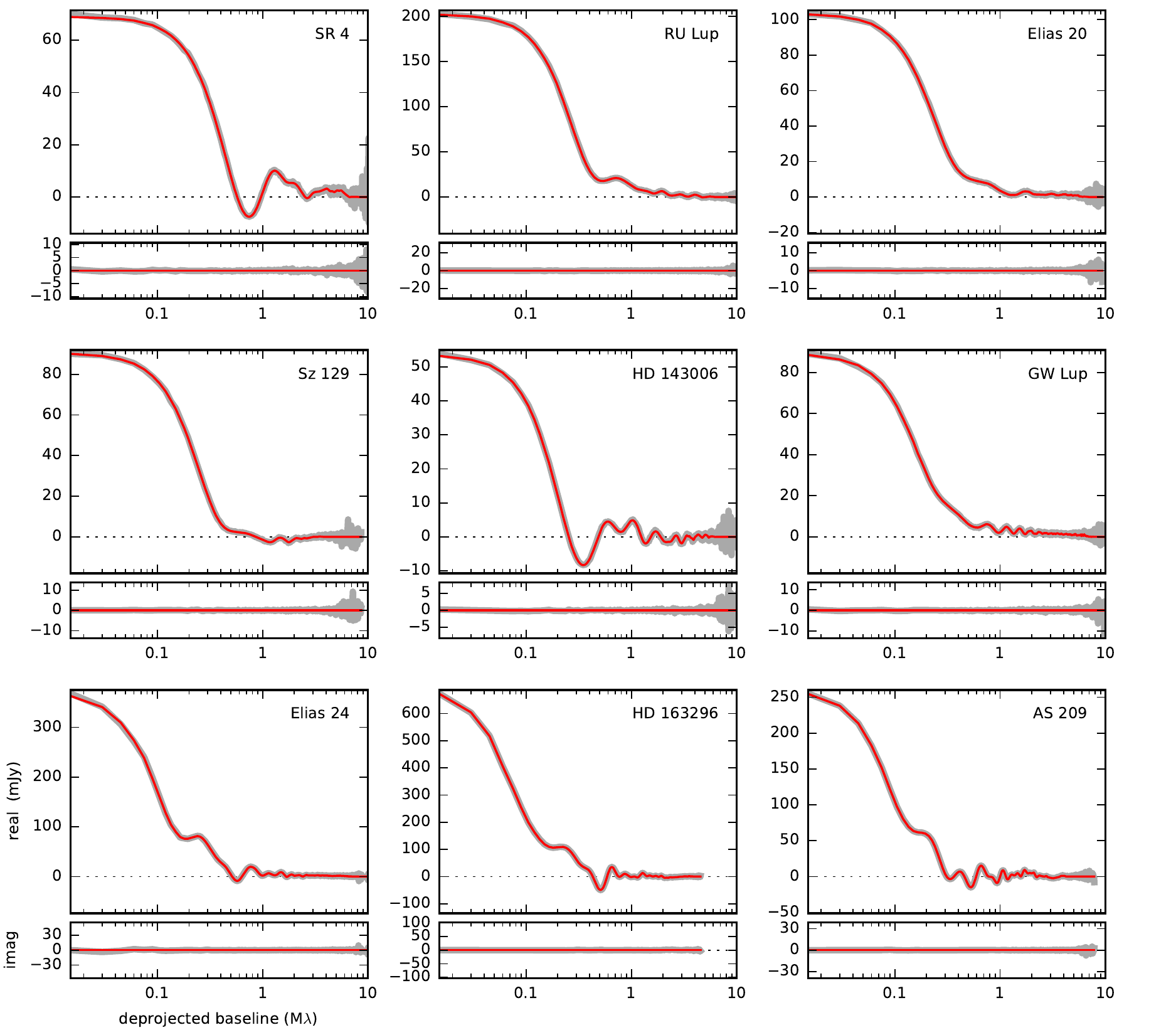}
\caption{The shifted, deprojected (see Table \ref{table:geom} for geometric parameters), and azimuthally-averaged (into 15 k$\lambda$-wide bins) visibilities as a function of baseline length for the data (gray) and the optimized {\tt frank} model (red).  Note that by definition the imaginary components of the models are zero for all spatial frequencies.  
\label{fig:visprofs}
}
\end{figure*}

\section{Model Radial Profiles} \label{app:SBprof}

Figure \ref{fig:Tbprofs} shows the model brightness profiles derived with {\tt frank} following the procedure outlined in Section \ref{sec:disk_removal}.  The gray bands mark the search zones for CPD emission, as described in Section \ref{sec:CPDs}, and the dotted lines denote the gap centers identified by \citet{dsharp2} (using a different approach).  There is good agreement with those results, though the {\tt frank} modeling hints at some annular features that are unresolved in the nominal DSHARP images (e.g., the gap at $\sim$30 mas in the SR 4 disk first identified by \citealt{jennings21}).

These profiles were used to crudely measure the centers, widths, and depths of the gaps of interest, as noted in Section \ref{sec:disk_removal}.  These estimates were made by visual comparison to a simplistic model of a background (local) power-law profile with a Gaussian depletion,
\begin{equation}
    \begin{aligned}
        T_b &= T_0 \, \frac{(r \, / \, 0\farcs1)^{-q}}{(1 + \Gamma)}, \,\,\,\, {\rm where} \\
        \Gamma &= (\delta_{\rm gap}{-}1) \exp{\left[-\frac{(r - r_{\rm gap})^2}{2 \sigma_{\rm gap}^2}\right]}.
    \end{aligned}
    \label{eq:gaps}
\end{equation}
The adopted power-law ($T_0$, $q$) and gap ($r_{\rm gap}$, $\sigma_{\rm gap}$, $\delta_{\rm gap}$) parameters are compiled in Table \ref{table:gaps}.  In the context of Equation (\ref{eq:gaps}), the depletion parameter $\delta_{\rm gap}$ is a multiplicative scale (amplitude); e.g., $\delta_{\rm gap} = 10$ implies relative depletion by an order of magnitude at $r_{\rm gap}$.  

Finally, to demonstrate the fit quality for the circumstellar disk emission in the native (Fourier) domain, Figure \ref{fig:visprofs} compares the deprojected, azimuthally-averaged visibilities with the corresponding {\tt frank} models.

\section{The CPD Model in the Fourier Domain} \label{app:CPD_model}

To quantify the sensitivity to CPD emission in the DSHARP data, we adopted the approach described in Section \ref{sec:CPDs} that characterizes the recovery rate of injected CPD signals.  In this framework, the mock CPD signal is computed in the Fourier domain and added to the observed (complex) `data' visibilities, ${\sf V_{\rm d}}({\sf u}, {\sf v})$, where ${\sf u}$, ${\sf v}$ are the Fourier spatial frequency coordinates (in wavelength units).  Those composite visibilities, ${\sf V_{\rm d}} + {\sf V_{\rm cpd}}$, are modeled with {\tt frank}, and the imaged residual visibilities are then searched for remnant CPD emission and compared with the known input parameters.

The CPD emission model is an offset point source with flux density $F_{\rm cpd}$ at ($r_{\rm cpd}$, $\varphi_{\rm cpd}$) in the disk-frame,
\begin{equation}
    I_{\rm cpd} = F_{\rm cpd} \, \delta(r{-}r_{\rm cpd}, \varphi {-}\varphi_{\rm cpd}),
\end{equation}
where $\delta$ is the Dirac $\delta$-function.  The Fourier transform of a point source at the origin is a constant (DC offset), in this case just $F_{\rm cpd}$.  But the offset position introduces an analytic oscillatory behavior.  The sky-frame coordinates of the mock CPD  ($x_s$, $y_s$) are given by Equation (\ref{eq:proj_geom}) for $z_{\rm s} = 0$.  For a given set of geometric parameters (Table \ref{table:geom}), the CPD visibilities can be expressed as
\begin{equation}
    {\sf V}_{\rm cpd}({\sf u}, {\sf v}) = F_{\rm cpd} \, e \,^{ 2 \pi i \left[ {\sf u}(x_s + \Delta x) + {\sf v}(y_s + \Delta y) \right] },
\end{equation}
where sky-projected terms are expressed in radians and in this case $i$ is the imaginary unit (not inclination).

\section{CPD Mass Constraints} \label{app:CPD_mass}

The framework we adopted to convert the derived flux constraints into limits on CPD masses follows closely the approach outlined by \citet{isella14,isella19}.  For convenience, we review the details here.

Physical models for the CPD continuum emission require assignments of the temperature and density structure.  We treat the CPD as geometrically flat and axisymmetric.  The CPD thermal structure can be approximated with the contributions of three mechanisms -- irradiation by the planet, irradiation by the star and local disk (around the gap), and accretion heating -- such that
\begin{equation}
    T^4 = T_{\rm irr, p}^4 + T_{\rm irr, \ast}^4 + T_{\rm acc}^4.
    \label{eq:CPD_heating}
\end{equation}
There are a lot of parameters and assumptions hidden in Equation (\ref{eq:CPD_heating}); each will be clarified below.

\begin{figure}[t!]
\includegraphics[width=\linewidth]{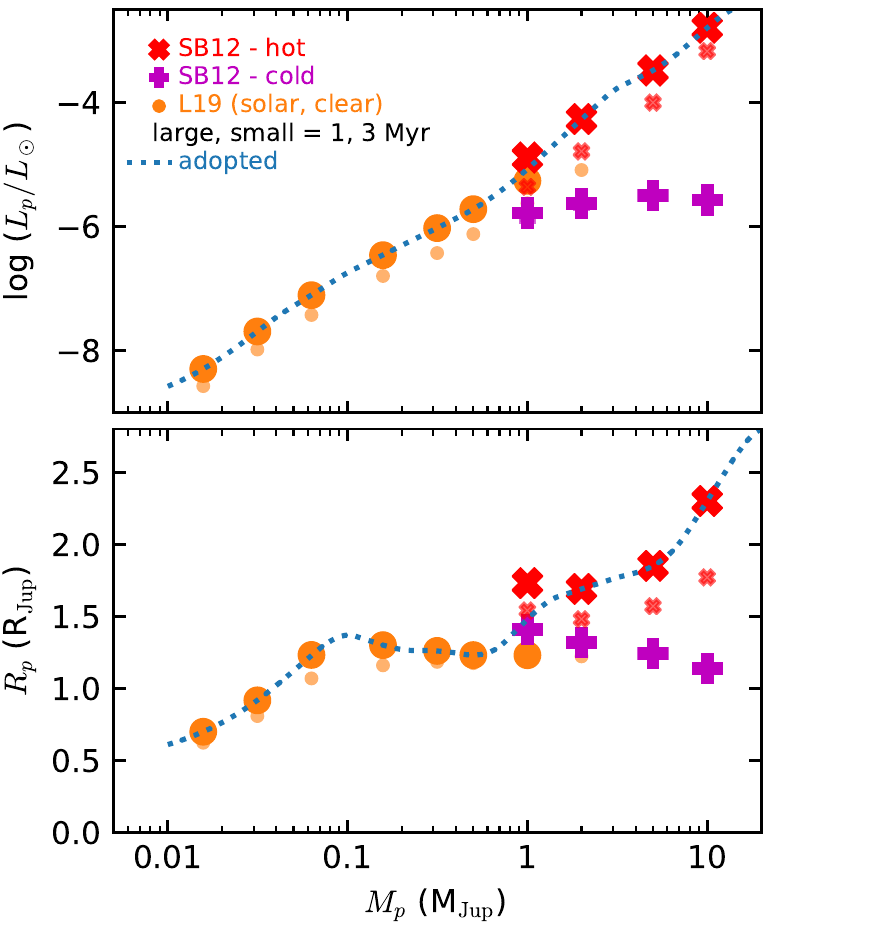}
\caption{The adopted variations of planet luminosity and radius as a function of mass used in CPD model calculations (blue, dotted curves), compared with predictions from detailed planetary evolution models at ages of 1 and 3 Myr (large and small symbols, respectively).  The orange circles show the solar metallicity predictions from \citet{linder19}, using the {\tt petitCODE} atmosphere modeling code and assuming cloud-free atmospheres.  The red {\sf X} and magenta $+$ markers denote the ``hot start" and "cold start" models from \citet{spiegel12} (their Figure 5).  The adopted curves are derived from spline interpolations of the \citet{linder19} 1 Myr models and the \citet{spiegel12} 1 Myr ``hot start" models at each $M_p$.
\label{fig:planet_evol}
}
\end{figure}

The heating contribution of irradiation by the ``host" planet can be approximated as
\begin{equation}
    T_{\rm irr, p}^4 = \frac{0.1 L_p}{4 \pi \sigma r^2},
\end{equation}
where $L_p$ denotes the planet luminosity, the factor 0.1 corresponds to the fraction of energy absorbed by a CPD with a vertical pressure scale height $\sim$10\%\ of the radius, $\sigma$ is the Stefan-Boltzmann constant, and $r$ is the radial coordinate in the CPD frame.  Since no planets associated with the disk gaps of interest here have been detected, we relied on theoretical models to estimate $L_p$ for any given $M_p$ (a free parameter).  The curve in the top panel of Figure \ref{fig:planet_evol} shows the adopted relationship, derived from an  interpolation of planetary evolution models (at a fixed age of 1 Myr) calculated by \citet{linder19} (solar metallicity models from the {\tt petitCODE} atmosphere modeling code) and \citet{spiegel12} (the ``hot start" models shown in their Figure 5). 

For models in ``low" accretion states (see below), this planetary irradiation heating dominates at least the inner part of the CPD.  The model planet luminosities decrease by a factor of 2--3 from 1 to 3 Myr, which amounts to only 20--30\%\ in $T_{\rm irr, p}$.  Since most of the emission comes from larger radii where other heating terms contribute, the corresponding decrease in the continuum emission (and thereby CPD mass estimates) is only $\sim$5--15\%.  For more massive planets, if we instead adopted the ``cold start" models from \citet{spiegel12}, the decrease in $L_p$ is more like a factor of 100.  While that decreases $T_{\rm irr, p}$ by a factor of $\sim$3, the emission decrease is much more muted ($\sim$30--40\%) because stellar/disk irradiation heating then dominates (particularly at larger $r$, where most of the emission originates).    

The irradiation heating from the central star and the local (circumstellar) disk is roughly constant in $r$, with
\begin{equation}
    T_{\rm irr, \ast}^4 = \frac{\phi L_\ast}{8 \pi \sigma r_{\rm gap}^2},
\end{equation}
where $L_\ast$ is the luminosity of the central star, $r_{\rm gap}$ is the planet/CPD location within the host disk (see Table \ref{table:gaps}), and $\phi$ is the flaring angle of the host disk.  We adopted the $L_\ast$ values catalogued by \citet{dsharp1} and the $\phi$ values used by \citet{dsharp2} (0.05 for the RU Lup disk, 0.02 otherwise).  If we generously allow that both $\phi$ and $L_\ast$ are uncertain by a factor of 2, that corresponds to a $\sim$40\%\ ambiguity in $T_{\rm irr, \ast}$.  For lower $M_p$ where $T_{\rm irr, \ast}$ dominates, this propagates almost directly into the CPD continuum luminosity; at larger $M_p$, the effect is considerably smaller (5--10\%).     

The viscous heating term can be approximated as
\begin{equation}
    T_{\rm acc}^4 = \frac{3 G M_p \dot{M}_p}{8 \pi \sigma r^3} \left(1 - \sqrt{\frac{R_p}{r}}\right)
\end{equation}
\citep[cf.,][]{dalessio98} where $\dot{M}_p$ is the planet accretion rate, $R_p$ is the planet radius (presumed equivalent to the inner edge of the CPD), and $G$ is the gravitational constant.  The adopted $R_p$ also depends on $M_p$ and an approximation of planetary evolution models, as shown in the bottom panel of Figure \ref{fig:planet_evol}.  We made the approximation that $\dot{M}_p = \eta M_p / t_p$, where $t_p$ is a characteristic timescale (assumed to be 1 Myr) and $\eta$ is an efficiency factor.  The ``low" (default) accretion state has $\eta = 1$, and a ``high" state has $\eta = 100$.  In the low state, irradiation heating dominates for all $M_p$; swapping $t_p$ to 3 Myr makes essentially no difference in the CPD fluxes.  But in the high state, viscous heating dominates in many cases (at least for $M_p \gtrsim 0.1$M$_{\rm Jup}$) and the output continuum fluxes are $\sim$2$\times$ higher.  There is considerable ambiguity associated with this heating term, since we do not yet understand the details of the CPD accretion process.  We consider the low and high state cases reasonable boundary conditions on the associated uncertainties (a factor of $\sim$two in the continuum emission levels).

The CPD density structure was described with a radial power-law for the surface densities, 
\begin{equation}
    \Sigma = \Sigma_0 \left(\frac{r}{R_{\rm cpd}}\right)^{-\gamma}
\end{equation}
defined for $R_p \le r \le R_{\rm cpd}$, where
\begin{equation}
    \Sigma_0 = (2-\gamma) \frac{M_{\rm cpd}}{2 \pi R_{\rm cpd}^\gamma} \left(R_{\rm cpd}^{2-\gamma} - R_p^{2-\gamma}\right)^{-1}.
\end{equation}
Here, $M_{\rm cpd}$ and $R_{\rm cpd}$ are the CPD mass and outer radius, respectively.  We set $\gamma = 0.75$, comparable to the analytical models of \citet{canup02}.  The flux differences associated with different $\gamma$ (e.g., from 0 to 1.5) are relatively small ($\sim$20\%), but depend in detail on the optical depths and heating terms for a given model.

The continuum flux from this CPD model is 
\begin{equation}
    F_{\rm cpd} = \frac{2 \pi \mu}{d^2} \int B_\nu(T) \left[1 - e^{-\tau_\nu/\mu} + \omega_\nu \mathcal{F} \right] \, r \, dr
    \label{eq:CPD_flux}
\end{equation}
\citep{miyake93,zhu19,sierra19}, where $\mu = \cos{i}$ is the direction cosine (with $i$ assumed equivalent to the inclination in Table \ref{table:geom}), $d$ is the distance to Earth (Table \ref{table:data}), and $B_\nu(T)$ the Planck function.  The dust grain emission properties are characterized with an absorption opacity $\kappa_\nu$ and albedo $\omega_\nu$.  Then, the optical depth is $\tau_\nu = \kappa_\nu (1 - \omega_\nu)^{-1} \Sigma$.  The scattering correction term in Equation (\ref{eq:CPD_flux}) is
\begin{equation}
    \mathcal{F} = \frac{ \displaystyle \left[ \frac{1{-}e^{-\sqrt{3} \epsilon_\nu \tau_\nu - \tau_\nu/\mu}}{\sqrt{3} \epsilon_\nu \mu + 1} + 
        \frac{e^{-\tau_\nu / \mu}{-}e^{-\sqrt{3} \epsilon_\nu \tau_\nu}}{\sqrt{3} \epsilon_\nu \mu - 1} \right] }{ \displaystyle (\epsilon_\nu{-}1)e^{-\sqrt{3} \epsilon_\nu \tau_\nu} - (\epsilon_\nu{+}1) },
\end{equation}
where $\epsilon_\nu = \sqrt{1 - \omega_\nu}$.  The main text explores the effect of albedo on the CPD flux (and mass) estimates for extreme boundary conditions.  Of course, the ambiguities associated with the optical properties of the particles that emit the mm continuum studied here are relevant and are expected to dominate the $M_{\rm cpd}$ uncertainties; more details on this general problem are discussed elsewhere \citep[e.g.,][]{dsharp5,andrews20}.    

\clearpage

\bibliography{references}

\end{document}